\documentclass[aps,prx,reprint]{revtex4-2}
\usepackage{blindtext}
\usepackage{graphicx}
\usepackage{amsmath}
\usepackage{amssymb}
\usepackage[english]{babel}
\usepackage{xcolor}
\usepackage{siunitx}
\usepackage[colorlinks=true, pdfstartview=FitV, linkcolor=blue,
citecolor=blue, urlcolor=blue]{hyperref}
\usepackage[capitalise]{cleveref}

\usepackage{float}

\makeatletter
\let\saved@includegraphics\includegraphics
\AtBeginDocument{\let\includegraphics\saved@includegraphics}
\renewenvironment*{figure}{\@float{figure}}{\end@float}

\makeatother

\makeatletter
\newcommand\footnoteref[1]{\protected@xdef\@thefnmark{\ref{#1}}\@footnotemark}
\makeatother	

\begin{document}

\title{Hole spin qubits in thin  curved quantum wells}

\author{Stefano Bosco}
\email{stefano.bosco@unibas.ch}
\author{Daniel Loss}

\affiliation{Department of Physics, University of Basel, Klingelbergstrasse 82, 4056 Basel, Switzerland}

\begin{abstract}
Hole spin qubits are frontrunner platforms for scalable quantum computers because of their large spin-orbit interaction which enables ultrafast all-electric qubit control at low power. The fastest spin qubits to date are defined in long quantum dots with two tight confinement directions, when the driving field is aligned to the smooth direction. However, in these systems the lifetime of the qubit is strongly limited by charge noise, a major issue in hole qubits. We propose here a different, scalable qubit design,  compatible with planar CMOS technology, where the hole is confined in a curved germanium quantum well surrounded by silicon. This design takes full advantage of the strong spin-orbit interaction of holes, and at the same time suppresses charge noise in a wide range of configurations, enabling highly coherent, ultrafast qubit gates. While here we focus on a Si/Ge/Si curved quantum well, our design is also applicable to different semiconductors.
Strikingly, these devices allow for ultrafast operations even in short quantum dots by a transversal electric field. This additional driving mechanism relaxes the demanding design constraints, and opens up a new  way to reliably interface spin qubits in a single quantum dot to microwave photons.
By considering state-of-the-art high-impedance resonators and realistic qubit designs, we estimate interaction strengths of a few hundreds of MHz, largely exceeding the decay rate of spins and photons. Reaching such a strong coupling regime in hole spin qubits will be a significant step towards high-fidelity entangling operations between distant qubits, as well as fast single-shot readout, and will pave the way towards the implementation of a large-scale semiconducting quantum processor.
\end{abstract}

\maketitle

\section{Introduction}

Spin qubits in hole quantum dots are frontrunner candidates to process quantum information and implement large-scale universal quantum computers~\cite{scappucci2020germanium,Gonzalez-Zalba2021,hendrickx2020fast,hendrickx2020four,Jirovec2021,maurand2016cmos,camenzind2021spin,piot2022single}.
The key advantages of holes stem from their reduced sensitivity to the noise caused by hyperfine interactions to defects with non-zero nuclear spin ~\cite{PhysRevB.78.155329,prechtel2016decoupling,warburton2013single,PhysRevLett.127.190501}, and from their strong effective spin-orbit interaction (SOI), which enables ultrafast and all-electric qubit control at low power~\cite{PhysRevLett.98.097202,froning2020ultrafast,Wang2022,watzinger2018germanium}.

The emergence of a large SOI in hole nanostructures is tightly linked to the design of the quantum dot, and it is maximized in systems where the hole is tightly confined in two directions  and electrically driven by a field aligned to the softer confinement~\cite{DRkloeffel1,DRkloeffel2,DRkloeffel3,bosco2021squeezed}.
 Rabi frequencies exceeding 400~MHz have been measured in germanium/silicon (Ge/Si) core/shell nanowires~\cite{froning2020ultrafast} and  Ge hut nanowires~\cite{Wang2022}. A major issue in these systems, however, is the large coupling to charge noise, that limits the coherence time  of the qubit to tens of nanoseconds.

Moreover, the strong SOI in long hole quantum dots is predicted to enable a strong transversal~\cite{DRkloeffel2} and longitudinal~\cite{bosco2022fully} coupling to high-impedance microwave resonators~\cite{PhysRevX.7.011030,PhysRevApplied.5.044004,Grunhaupt2019,Maleeva2018,PhysRevApplied.11.044014,PhysRevLett.121.117001}, where the strength of the interaction exceeds the coherence time of the qubits and the photons.
Reaching the strong coupling regime in hole quantum dots will enable long-range connectivity of  distant qubits~\cite{VandersypenInterfacingspinqubits2017} as well as quantum error correcting architectures~\cite{PhysRevLett.118.147701}, and will be a big step towards scaling up spin-based quantum computers.
In contrast to state-of-the-art experiments, where multiple quantum dots encode a single qubit~\cite{Landig2018,mi2018coherent,doi:10.1126/science.aaa3786,harvey2021circuit,PhysRevB.100.125430,bottcher2021parametric}, hole spin qubits only need a single quantum dot~\cite{DRkloeffel2,bosco2022fully,PhysRevB.102.205412,michal2022tunable}, significantly diminishing the complexity of the architecture. However, to enhance the spin-photon interactions, the zero-point field of the photon needs to be aligned to the long direction of the dot~\cite{DRkloeffel2,bosco2022fully}, and thus the plunger electrode capacitively coupling the dot and the resonator has to be misaligned from center of the dot, reducing the geometric lever arm to this gate and limiting the maximal spin-resonator coupling strength.\\

In this work, we propose a different type of hole spin qubit that is defined in a thin curved quantum well (CQW).
This architecture not only benefits from the large SOI of hole quantum dots, which can  even be reached at lower values of electric field, but  it also can be designed to be free of charge noise. In striking contrast to alternative proposals to reduce the charge noise~\cite{bosco2020hole, Wang2021,Malcok2022}, in CQWs charge noise is suppressed for a wide range of electric fields and not only at fine-tuned points in parameter space, providing a critical technological advantage compared to competing architectures. This enhancement could push spin-based quantum information processing towards new speed and coherence standards.

The smallest coupling to charge noise occurs when the magnetic field is aligned to the well, where the effective Zeeman energy and the $g$-factor is widely tunable  by external electric fields and by engineering the strain of the well. Strikingly, in an annular CQW, the maximal $g$-factor can reach rather large values,  in analogy to topological insulator nanowires~\cite{PhysRevB.104.165405,legg2021giant}. For this reason, CQWs can be  effective architectures also in the search of exotic topological particles, such as Majorana fermions~\cite{doi:10.1063/5.0055997}, where the low value of the $g$-factor along the hole nanowire is a critical issue~\cite{PhysRevB.90.195421}.

Moreover, because of the large dipole moment of holes confined in CQWs, even  spin qubits  in short quantum dots can be driven ultrafast at low power by electric fields perpendicular to the smooth confinement direction. This mechanism not only relaxes the stringent technological constraints on the design of ultrafast spin qubits, but it also offers a different way to interface spin qubits in single quantum dots to microwave photons in high-impedance resonators, thus pushing spin-photon hybrid architectures towards new speed and coherence standards, and paving the way towards the implementation of a large-scale quantum processor.\\

This work is organized as follows.
In Sec.~\ref{sec:theory}, we present the system and we introduce the theoretical model used to describe it, including an analysis of the strain, a crucial ingredient in hole nanostructures~\cite{PhysRevB.90.115419,Niquet2012,PhysRevB.103.245304}. We discuss two different setups: an annular CQWs, where a semiconducting shell fully covers an inner core~\cite{Lauhon2002}, and a CQW grown on top of a planar substrate. The latter architecture, in particular, is  compatible to current CMOS processes, and it holds particular promise for scaling up quantum computers.  
As the complete theoretical model is rather complicated, to gain valuable insights into the response of the system, in Sec.~\ref{sec:Effective-theory}, we derive an effective low energy theory for the CQW, and discuss its key features in the presence of electric and magnetic fields. The effective model introduced in this section describes with a reasonable accuracy a wide range of devices, also when additional valence bands as well as lattice and cross-section anisotropies are included. These effects are addressed in Apps.~\ref{app:SOHs} and~\ref{sec:deviation_LK}, respectively.
In Sec.~\ref{sec:SQD} and Sec.~\ref{sec:Elongated-QDs}, we analyze in detail hole spin qubits in these systems and we discuss  qubits in quantum dots that are short and long compared to the radius of the well, respectively. We highlight the advantages of these qubits compared to alternative designs and we also examine the differences between annular and planar CQWs.
In Sec.~\ref{sec:spin-photon}, we analyze the coupling between these qubits and photons in microwave resonators. We estimate that the interaction strength in state-of-the-art devices can exceed a few hundreds of MHz, much larger than the decay rate of qubits and photons, thus enabling a strong qubit-photon coupling, with far-reaching consequences for spin-based quantum computing.

\section{Holes in curved quantum wells}
\label{sec:theory}
\subsection{Setup}
\label{sec:Setup}

\begin{figure}
\centering
\includegraphics[width=0.45\textwidth]{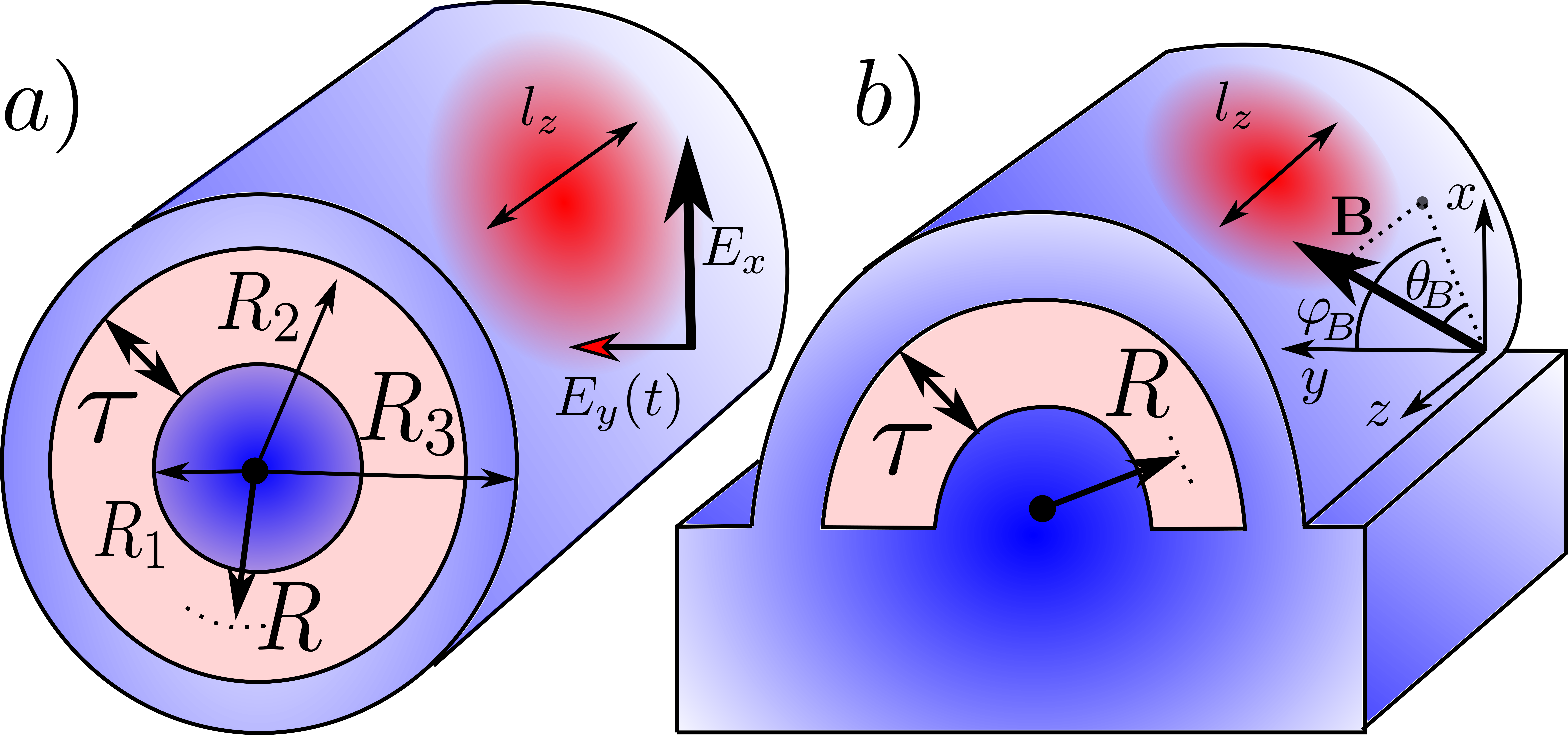}
\caption{\label{fig:sketch} Sketch of a hole spin qubit in a curved quantum well. The hole wavefunction is confined in the pink region in the $(x,y)$-plane, and it extends for a harmonic length $l_z$ in the $z$-direction. External electric and magnetic fields are also indicated. The blue regions highlight the interface with a different material. In particular, an experimentally relevant example comprises a germanium thin quantum well (pink) surrounded by silicon (blue). In a) we show an annular CQW, where a thin semiconducting shell fully surrounds an inner core. In b) we show a CQW in a planar setup, where the semiconducting shell is grown on top of a planar substrate. }
\end{figure}

In this work, we analyze the CQWs sketched in Fig.~\ref{fig:sketch}. 
We examine setups where the hole wavefunction is confined  in an annular CQW (pink)  sandwiched between a  core of radius $R_1$ and an outer shell (blue) extending from a radius $R_2$ to $R_3$. We introduce the thickness $\tau=R_2-R_1$ and the average radius $R=(R_1+R_2)/2$ of the quantum well.
The  well extends along the $z$-direction and the $x=r\cos{\varphi}$ and $y=r\sin(\varphi)$ directions define the cross-section; $r$ and $\varphi$ are polar coordinates. 

The  spin qubit is  defined by confining the hole in a quantum dot along  $z$ and by further applying a magnetic field $\textbf{B}$ that splits the spin states. The quantum dot can be defined electrostatically by metallic gates\cite{froning2020ultrafast,hendrickx2020four}, or by growing thin insulating regions in the quantum well~\cite{Jia2019}. While the latter approach could provide a better control over size and shape of the dot, for simplicity, we focus here on the former case~\footnote{We note that in the annular CQW of Fig.~\ref{fig:sketch}a), the technologically challenging gate-all-around technology can give an excellent control over the quantum dot dimension and shape.}. 

For the most part of our analysis, we explicitly study annular CQWs where a thin semiconducting shell  fully surrounds an inner core, as shown in Fig.~\ref{fig:sketch}a). This structure is compatible with current technology~\cite{Lauhon2002}. However, our theory describes well also architectures comprising planar CQWs grown on a planar substrate, as shown in Fig.\ref{fig:sketch}b), as well as CQWs with less symmetric cross-sections, e.g. square or hexagonal.
In particular, planar architectures grown on top of a silicon substrate hold particular promise to scale up quantum processors because of their compatibility with CMOS technology~\cite{Veldhorst2017}. 

In this context, another material that is attracting much attention is strained Ge, that present large SOI, small effective mass, and can be grown epitaxially with high purity on top of Si~\cite{scappucci2020germanium}.
Moreover, in nanowires the mismatch of the energy bands between Si and Ge and the alignment of their Fermi energies ensures that even without external gating, the charge carriers are holes confined in Ge~\cite{scappucci2020germanium}.
For these reasons, although our theory is applicable to a wide range of semiconducting materials, we restrict ourselves to the analysis of a Ge CQW (where the hole is confined) surrounded by Si. The confinement of the holes in the shell instead than in the core has far-reaching consequences on the resulting spin qubits, and sets our setup apart from current Ge/Si core/shell and hut nanowires~\cite{froning2020ultrafast,PhysRevResearch.3.013081, Wang2022}.

We now discuss in detail the key ingredients required to define these qubits, including a description of the physics of the valence band of semiconductors, of the electric and magnetic fields, and of strain. 

\subsection{Theoretical model}
\label{sec:Model}

The physics of  hole nanostructures is accurately described by the Hamiltonian
\begin{equation}
\label{eq:Hamiltonian}
H=H_\text{LK}+V_\text{C}(r,z) - e \textbf{E}\cdot \textbf{r} + H_\textbf{B}+ H_\text{BP} \ .
\end{equation}
The kinetic energy of the holes is modelled by the isotropic Luttinger-Kohn (LK) Hamiltonian~\cite{WinklerSpinOrbitCoupling2003}
\begin{equation}
\label{eq:LK-Hamiltonian}
H_\text{LK}=\left(\gamma_1+\frac{5}{2}\gamma_s\right)\frac{p^2}{2m} -\frac{\gamma_s}{m}(\textbf{p}\cdot \textbf{J})^2 \ ,
\end{equation}
where $\gamma_1$ and $\gamma_s$ are material-dependent LK parameters parametrizing the mixture of heavy holes (HHs) and light holes (LHs) at the top of the valence band of cubic semiconductors, $\textbf{p}=-i\hbar\nabla$ is the canonical momentum [$p^2=-\hbar^2\nabla^2$] and $\textbf{J}=(J_x,J_y,J_z)$ is the vector of spin-3/2 matrices. In particular, in Ge  $\gamma_1\approx 13.35$ and $\gamma_s\approx 4.96$~\cite{WinklerSpinOrbitCoupling2003}. Deviations from this model, including the contributions of the additional valence band and of cubic anisotropies are addressed in App.~\ref{app:SOHs} and~\ref{sec:deviation_LK}, respectively.
The quantum dot is defined by the confinement potential $V_\text{C}(r,z) =V_r(r)+\hbar\omega_z z^2/2l_z^2$, comprising an abrupt potential $V_r(r)$ in the radial direction modelling the boundary of the CQW and a smooth harmonic potential parametrized by a characteristic length $l_z$ and frequency $\omega_z=\hbar\gamma_1/ml_z^2$. 

We also include an electric field $\textbf{E}$ produced by the gates and a magnetic field $\textbf{B}$, which couples to the spin of the hole by the Hamiltonian 
\begin{equation}
\frac{H_\textbf{B}}{2\mu_B}=\kappa \textbf{B}\cdot \textbf{J}+\frac{2\gamma_s}{\hbar}\left[\left(\frac{\gamma_1}{\gamma_s}+\frac{5}{2}\right)\left\{\textbf{A}, \textbf{p}   \right\}- 2\left\{\textbf{A}\cdot\textbf{J}, \textbf{p}\cdot \textbf{J}   \right\}\right] \ ,
\end{equation}
where $\{A,B\}=(AB+BA)/2$ and $\mu_B$ is the Bohr magneton.
Here, $H_\textbf{B}$ includes the Zeeman field and the orbital magnetic field effects coming from the Peierls substitutions $\textbf{p}\to \pmb{ \pi}=\textbf{p}+e\textbf{A}$, with $\textbf{A}=(zB_y-yB_z/2,-zB_x+xB_z/2,0)$. We neglect small corrections $\mathcal{O}(\textbf{B}^2)$ and anisotropies $\propto J_i^3$ of the Zeeman interactions.

\subsection{Strain in a Ge curved quantum well}
\label{sec:strain}

\begin{figure}
\centering
\includegraphics[width=0.45\textwidth]{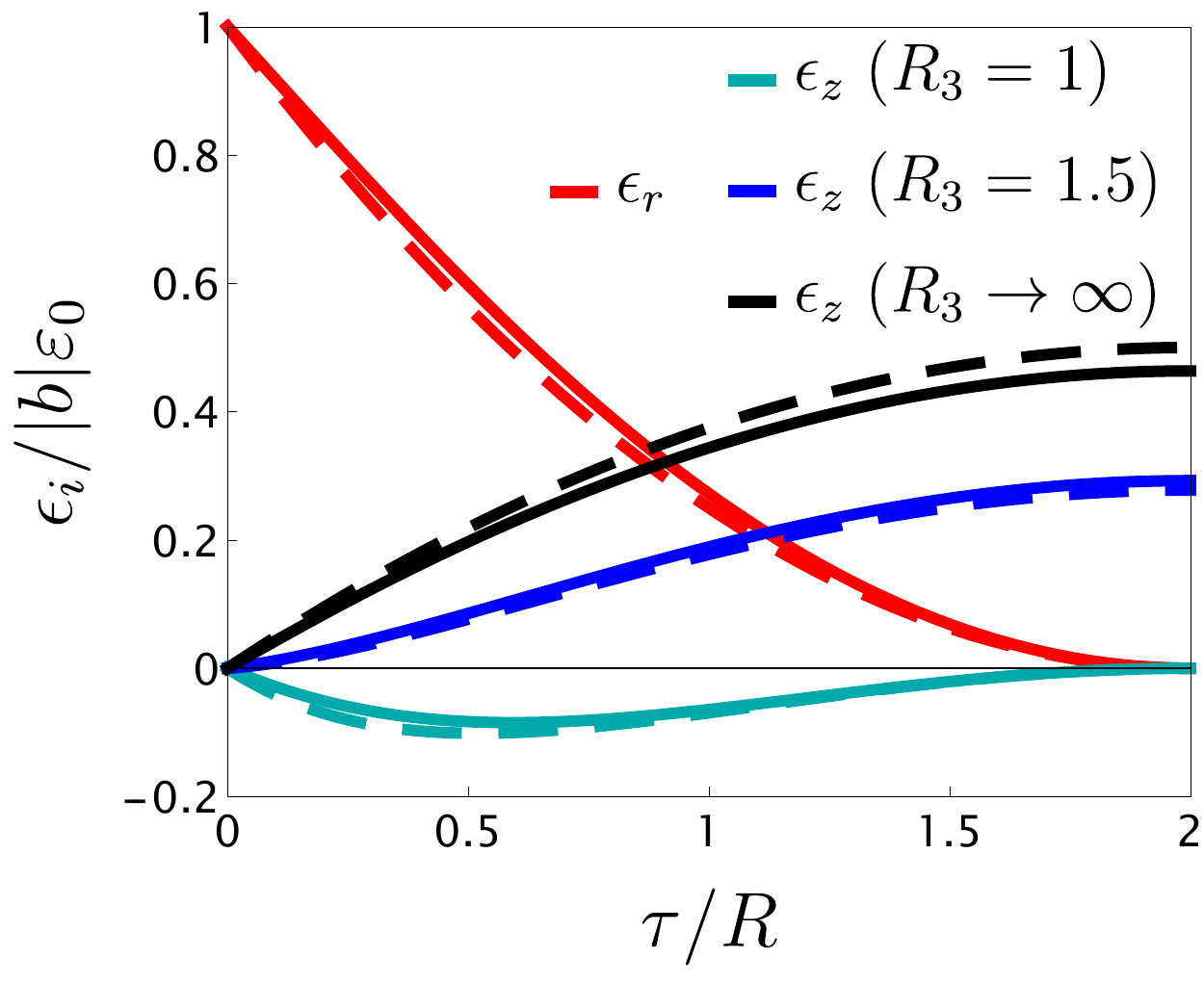}
\caption{\label{fig:strain} Strain energy of a  Ge CQW as a function of thickness $\tau$ of the well. With dashed lines we show the approximated results in Eq.~\eqref{eq_strain_pars}, while with solid lines we show a more general result, including differences in the elastic constants of Si and Ge, as discussed in App.~\ref{app:strain}. The radial strain $\epsilon_r$ is to good approximation independent of the outer Si shell, while the longitudinal strain $\epsilon_z$ depends strongly on the outer shell thickness $R_3$, measured with respect to $R_2$, see Fig.~\ref{fig:sketch}.
For Ge CQWs, the radial strain dominates, while the longitudinal strain plays a significant role in thicker Ge wells, especially when covered by a thick outer Si layer.  }
\end{figure}

Another key feature to understand the behaviour of hole spin qubits is the strain
that strongly hybridizes HHs and LHs~\cite{PhysRevB.90.115419,Niquet2012,PhysRevB.103.245304}.
We now restrict ourselves to the analysis of the technologically relevant scenario where a thin Ge CQW is surrounded by Si.
In this case, the strain is mainly determined by the mismatch of lattice constants of Si and Ge, $a_\text{Si}=0.543$~nm and $a_\text{Ge}=0.566$~nm, respectively. 

In the annular Ge CQW sketched in Fig.~\ref{fig:sketch}a), the strain is strikingly different than in other devices~\cite{PhysRevB.90.115419,doi:10.1002/adma.201906523,PhysRevB.103.125201,bosco2021squeezed}, and it is accurately modelled by the Bir-Pikus (BP) Hamiltonian
\begin{equation}
\label{eq:strain}
H_\text{BP}=J_z^2\epsilon_z- J_r^2\epsilon_r \ ,
\end{equation}
where $J_r=J_x\cos(\varphi)+J_y\sin(\varphi)$ is the spin-3/2 matrix aligned to the radial direction.
The strain energies $\epsilon_{r}$ and $\epsilon_z$ can be approximated as
\begin{subequations}
\label{eq_strain_pars}
\begin{align}
\epsilon_r &  \approx \left(1-\frac{\tau}{2R}\right)^2|b| \varepsilon_0 >0\ , \\
\epsilon_z& \approx
\left(\frac{1}{2}-\frac{\tau}{8R}-\frac{R^2}{R_3^2}\right) \frac{\tau}{R}|b|\varepsilon_0 \ .
\end{align}
\end{subequations}
where $\varepsilon_0\approx 1.6 \varepsilon_\parallel$ is the typical strain in planar heterostructure~\cite{doi:10.1063/1.1601686}, $\varepsilon_\parallel=(a_\text{Ge}-a_\text{Si})/a_\text{Ge}\approx 4\%$ is the relative mismatch of lattice constants of Si and Ge, and $b=-2.2$~eV~\cite{WinklerSpinOrbitCoupling2003}. 
If Ge is grown on pure Si, the typical strain energy is $|b|\varepsilon_0\approx 140$~meV. However, if a Si$_{1-x}$Ge$_x$ compound~\cite{scappucci2020germanium} substitutes pure Si in the core and outer shell, the  strain  decreases as $\varepsilon_\parallel\to (1-x)\varepsilon_\parallel$. 
The dependence of these quantities on the thickness $\tau$ of the Ge well  is shown in Fig.~\ref{fig:strain}. By comparing  to the results obtained by a more general analysis analogous to Ref.~\cite{PhysRevB.90.115419}, we observe that the simple expressions in Eq.~\eqref{eq_strain_pars} accurately describe the system.
A more detailed derivation of Eqs.~\eqref{eq:strain} and~\eqref{eq_strain_pars}, also including a discussion on inhomogeneous strain, is provided in App.~\ref{app:strain}.

From Eq.~\eqref{eq:strain}, we observe that the strain energy can be decomposed into two different components, parametrized by the competing energies $\epsilon_{r,z}$. We emphasize that these energies can be designed independently because they depend on different design parameters of the cross-section. In particular,  the energy $\epsilon_z$ favours holes with the quantization axis aligned to the $z$-direction, and it strongly depends on thickness $\tau$ of the Ge well and on the radius $R_3$ of the outer Si shell. When $R_3$ is sufficiently large (small), then $\epsilon_z>0$  ($\epsilon_z<0$) and LHs (HHs) have a lower energy. When $\epsilon_z>0$, this terms is qualitatively analogous to the typical strain in Si/Ge core/shell nanowires~\cite{PhysRevB.90.115419}, where the BP Hamiltonian is~\footnote{The strain of a Ge/Si core/shell nanowire with inner radius $R_1$ and outer radius $R_2$ is straightforwardly related to the strain in the  thin Ge CQW [Eq.~\eqref{eq_strain_pars}] by the substitutions $\tau/R\to 2$ and $R/R_3\to R_1/2R_2$.}
\begin{equation}
H_\text{BP}^\text{c/s}\approx J_z^2  \frac{|b|\varepsilon_0}{2}\left(1- \frac{R_1^2}{R_2^2}\right) \ .
\end{equation}

In contrast, the energy $\epsilon_r$  favours HHs in the radial direction, and to good approximation it is independent of the presence of the outer Si shell. This type of strain is not present in usual Ge/Si core/shell nanowires, but it emerges in planar heterostructures~\cite{PhysRevB.103.125201,Wang2021,bosco2021squeezed} and hut wires~\cite{doi:10.1002/adma.201906523}, where the ground states are expected to be HHs with quantization axis aligned to the strong confinement direction. When $\tau\ll R$, we recover this expected limit and 
\begin{equation}
H_\text{BP}^\text{PH}=-J_r^2 |b|\varepsilon_0 \ .
\end{equation}
From the Fig.~\ref{fig:strain}, we observe that while $\epsilon_z$ is dominant at large values of the ratio $\tau/R$, in thin Ge CQWs $\epsilon_r$ is the dominant contribution and the ground state comprises to good approximation radial HHs.  

In the following, we restrict ourselves to the analysis of devices with thin Ge CQWs  and a thick outer Si layer, such that $\epsilon_r>\epsilon_z>0$.

\subsection{Hamiltonian in cylindrical coordinates }

\label{sec:frame}
In a thin Ge CQW, the radial confinement energy
 \begin{equation}
\epsilon_c=\frac{\hbar^2\pi^2\gamma_1}{m\tau^2}\approx 100 \times \left(\frac{10~\text{nm}}{\tau}\right)^2~\text{meV}
\end{equation}
is large compared to the quantization energy $\hbar\omega_\varphi=\hbar^2\gamma_1/mR^2$ of the total angular momentum and to the confinement energy $\hbar\omega_z=\hbar^2\gamma_1/ml_z^2$ along the quantum well, both in the meV range. In this case, where $\tau\ll R, l_z$, the dynamics of the confined holes is well-described by an effective low-energy theory where the radial degrees of freedom are frozen. 

However, because the radial strain $\epsilon_r$ is comparable to the radial confinement, i.e. $\epsilon_c\sim \epsilon_r\sim 100$~meV, an accurate low-energy model of the system needs to account exactly for $\epsilon_r$. For this reason, we rotate the Hamiltonian in Eq.~\eqref{eq:Hamiltonian} to a cylindrical coordinate system where the spin quantization axis is aligned to the radial direction and the radial strain $-\epsilon_r J_r^2$ is diagonal.
This rotation is generated by the angular-dependent unitary operator $U=e^{-i J_3 \varphi} e^{-i J_2 \pi/2}$, comprising a first rotation that aligns the spin matrices $J_x$ and $J_y$ to the radial and angular directions, respectively, and a second transformation that aligns the spin quantization axis to the radial direction, i.e. $J_r\to J_3$ and $J_z\to -J_1$.
To avoid confusion, in the new frame we label the spin-3/2 matrices by 1,2,3 instead of $x,y,z$. 

The effect of $U$ on the most relevant operators is
\begin{subequations}
\label{eq:rot_polar}
\begin{align}
\label{eq:mom-rot}
U^\dagger \textbf{p} U & = \textbf{p}+ \hbar \textbf{e}_\varphi  J_1 \ , \\
\label{eq:J-rot}
U^\dagger \textbf{J} U &= \textbf{e}_r J_3 + \textbf{e}_\varphi J_2 - \textbf{e}_z J_1 \ .
\end{align}
\end{subequations}
We note that in the product of the rotated operators, some care must be taken because $\textbf{e}_{r,\varphi}$ are unit vectors in cylindrical coordinates that depend on $\varphi$, and do not commute with $p_\varphi=-i\hbar\partial_\varphi$. For this reason, we report in App.~\ref{sec:LK-BP-cyl} the explicit expressions of the LK and BP Hamiltonians in this coordinate system.

Importantly, from Eq.~\eqref{eq:rot_polar} it follows that the total angular momentum $F_z=p_\varphi+\hbar J_z$ in the original coordinate system transforms as
\begin{equation}
U^\dagger F_z U  = p_\varphi \ ,
\end{equation}
in the new frame. Consequently, the physical eigensolutions of the Hamiltonian in the rotated frame are antiperiodic in $\varphi$ and $p_\varphi$ is quantized in units of $(2l-1)\hbar/2$, with $l\in\mathbb{Z}$.  

\section{Effective low-energy theory}
\label{sec:Effective-theory}
We first derive an effective model describing the band dispersion in the absence of external fields ($\textbf{E}=\textbf{B}=0$) and then generalize our results to account for finite values of $\textbf{E}$ and $\textbf{B}$.
In the  frame introduced in Sec.~\ref{sec:frame} and when $\textbf{E}=\textbf{B}=0$, $p_\varphi$ and $p_z$ are good quantum numbers of the total isotropic Hamiltonian $H$ in Eq.~\eqref{eq:Hamiltonian}. To construct an effective Hamiltonian that acts only on these two degrees of freedom, we trace out the radial direction by projecting the rotated $H$ onto the basis states
\begin{equation}
\label{eq:basis-states}
\psi_n(r)=\sqrt{\frac{2}{\tau r}}\sin\left[\frac{\pi n}{\tau} \left(r-R-\frac{\tau}{2}\right)\right] \ , \ n\in \mathbb{N} \ , 
\end{equation}
satisfying hard-wall boundary conditions at $r=R\pm \tau/2$.
These functions are the eigenstates of the operator $\gamma_1 p_r^2/m$ to eigenvalues $\epsilon_c n^2$, with $p_r=-i\hbar(\partial_r+1/2r)$ being the hermitian radial momentum, and they provide a complete basis for the radial degree of freedom.

\begin{figure}
\centering
\includegraphics[width=0.5\textwidth]{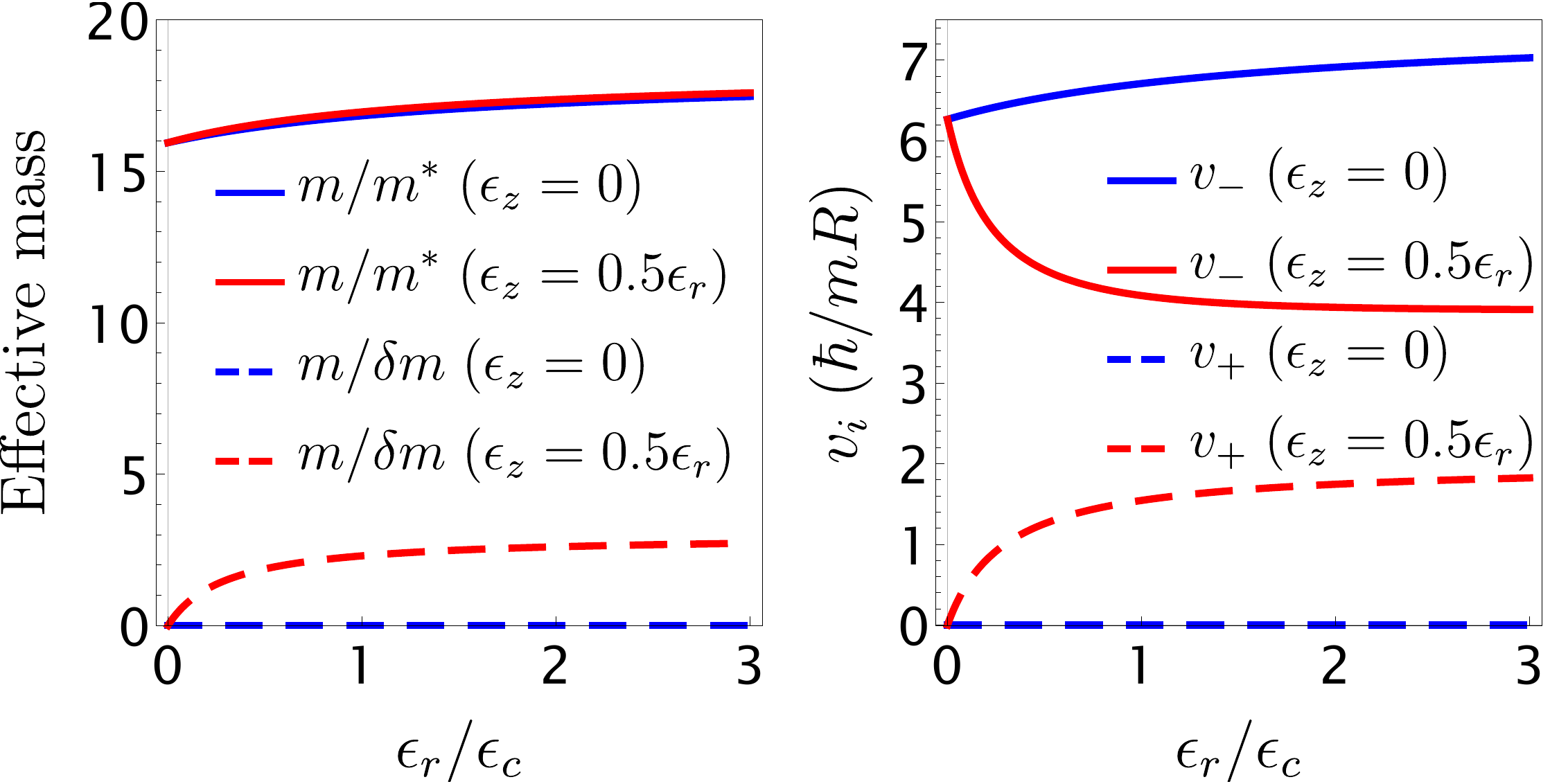}
\caption{\label{fig:pars} Parameters of the effective theory in Eq.~\eqref{eq:effective_H} of a thin Ge CQW. We show with blue and red lines the inverse effective mass and the spin-orbit velocities, see Eq.~\eqref{eq:pars},  obtained at $\epsilon_z=0$ and at $\epsilon_z=0.5\epsilon_r$, respectively. While the effective mass is not strongly modified by strain, the SOI can acquire a large strain dependence, and it significantly varies as a function of the longitudinal strain $\epsilon_z$.   }
\end{figure}

The quasi-degenerate ground state of the system comprises two HH Kramers partners with quantum number $n=1$, energetically separated from the first (HH) excited states by the energy $3\epsilon_c (1-2\gamma_s/\gamma_1)/2$, and from the second (LH) excited state by $2\gamma_s\epsilon_c/\gamma_1+2\epsilon_r+\epsilon_z$.
By considering only the first two radial eigenstates with $n=1$ and $n=2$, and by a second order Schrieffer-Wolff transformation~\cite{WinklerSpinOrbitCoupling2003,BRAVYI20112793} we find that the dynamics of the ground state is captured by the quadratic Hamiltonian 
\begin{equation}
\label{eq:effective_H}
H_\text{GS}=\frac{p_+p_-}{2m^*} -\frac{p_+^2+p_-^2}{4\delta m}+\sigma_+ \left( v_- p_--v_+  p_+ \right) + \text{h.c.}  \ ,
\end{equation}
where $p_\pm=\frac{p_\varphi}{R}\pm i p_z$ and $\text{h.c.}$ means hermitian conjugate.
The parameters of the effective theory to second order in $\tau/R$ are given by
\begin{subequations}
\label{eq:pars}
\begin{align}
\frac{1}{m^*}&\approx \frac{1}{m}\left(\gamma _1+\gamma _s -3 \tilde{\gamma}\right)\ , \\
v_-&\approx \frac{3}{2}\frac{\hbar}{m R}\left[\left(\gamma_s-\tilde{\gamma}\right)  -\frac{1}{2}(\gamma_1+2\gamma_s)\tilde{\epsilon}_z \right]\ , \\
\frac{1}{\delta m}&\approx \frac{3}{m} \gamma_s \tilde{\epsilon}_z \ , \ \ \text{and} \ \ \ v_+ \approx \frac{3}{4}\frac{\hbar}{m R}\gamma_1\tilde{\epsilon}_z \ ;
\end{align}
\end{subequations}
we introduce here the dimensionless quantities
\begin{subequations}
\begin{align}
\tilde{\gamma}&=\frac{256}{9 \pi^2}\frac{ \gamma _s }{10 + \gamma _1 \left(3 \epsilon _c+4 \epsilon _r+2 \epsilon _z\right)/\gamma_s \epsilon _c } \ , \\
\tilde{\epsilon}_z&= \frac{ \epsilon _z}{\epsilon _z+2 \epsilon _r+2 \gamma _s\epsilon _c/\gamma _1 } \ .
\end{align}
\end{subequations}

While $\tilde{\gamma}$ only quantitatively modifies the parameters of the effective Hamiltonian, the longitudinal strain $\tilde{\epsilon}_z$ introduces the qualitatively different terms $v_+$ and $ \delta m$, that modify the SOI and the effective mass along the quantum well and in the angular direction. The dependence of these parameters on strain is shown in Fig.~\ref{fig:pars}. We observe that while the effective mass is not strongly affected by $\delta m$, the effective SOI can be largely modified by $\epsilon_z$, and $v_+$ can become comparable to $v_-$. 

The effective theory  in Eq.~\eqref{eq:effective_H} can be also generalized to include the high energy valence band that is separated in energy from the HH-LH subspace by an energy $\Delta_\text{S}\approx 300$~meV. These high energy states do not alter qualitatively the physics described here. In fact, in this case, Eq.~\eqref{eq:effective_H} is still valid, but the effective parameters in Eq.~\eqref{eq:pars} acquire corrections that scale as $\Delta_\text{S}/\epsilon_c$. The effect of the additional valence band is discussed in detail in App.~\ref{app:SOHs}, and in particular,  the generalized version of Eq.~\eqref{eq:pars} is given in  Eq.~\eqref{eq:pars_SOH}.  \\

\begin{figure}
\centering
\includegraphics[width=0.5\textwidth]{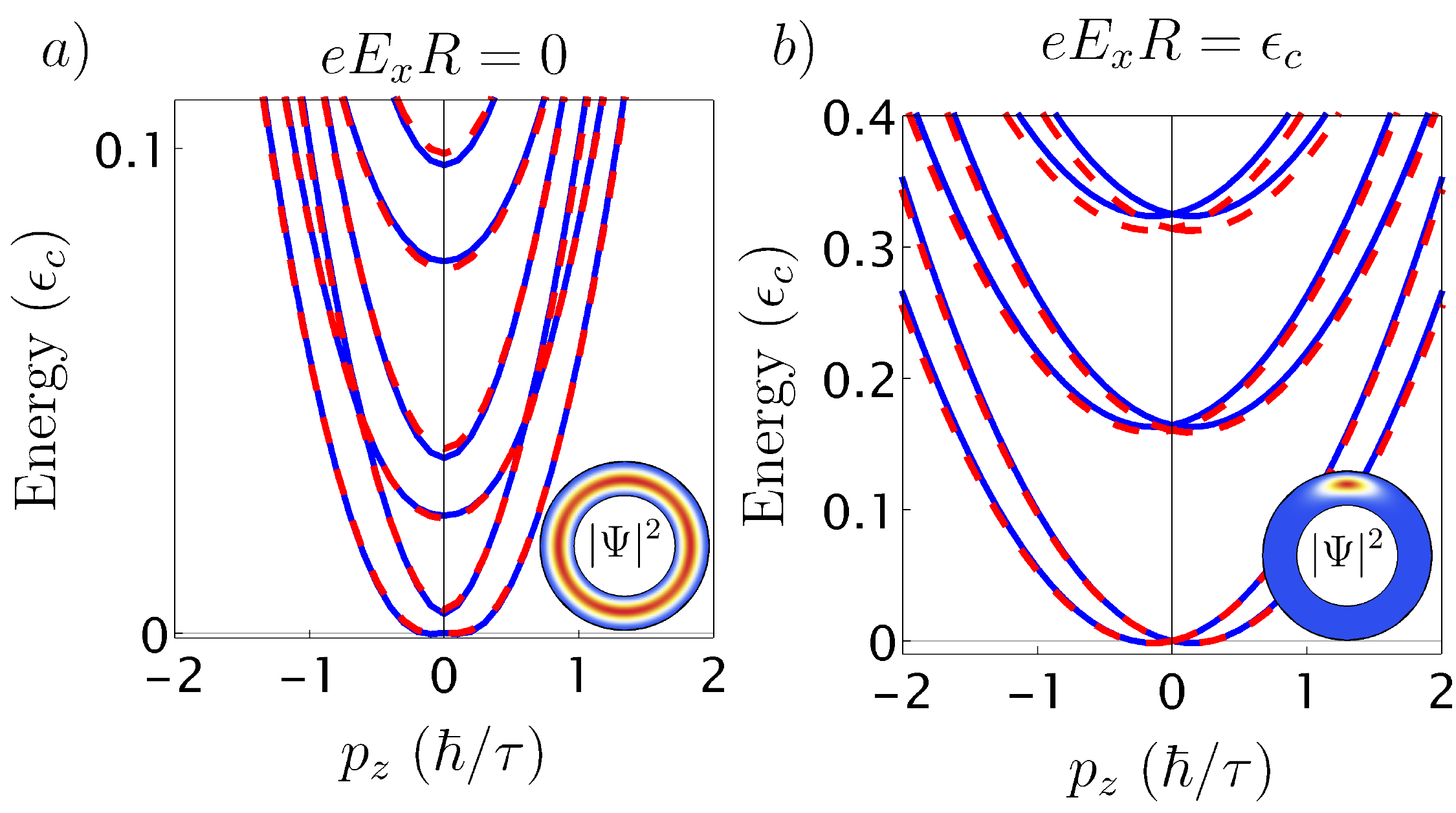}
\caption{\label{fig:band} Energy dispersion of a strained thin Ge CQW. We compare the dispersion calculated from the effective theory in Eq.~\eqref{eq:effective_H} (blue lines) and by numerically diagonalizing the Hamiltonian $H$ in Eq.~\eqref{eq:Hamiltonian} (red dashed lines). In a) and b), we show results obtained at $E_x=0$ and $E_x=\epsilon_c/eR$. We consider $\textbf{B}=0$, $\tau=R/2$, $\epsilon_r=2\epsilon_z=3\epsilon_c$. In the insets at the bottom right of the figures, we show the hole density of the groundstate at $p_z=0$ in the lab frame.  
At $p_z=0$, in a) the ground and first excited states are spilt by the small SOI gap $\Delta$ [Eq.~\eqref{eq:delta}]. In contrast, in b) the gap $\hbar\omega_E$ [Eq.~\eqref{eq:omegaE}]  is large and $E_x$ pins the hole to the top of the well.  }
\end{figure}

In Fig.~\ref{fig:band}a) we compare the energy dispersion of the lowest energy levels as a function of $p_z$ derived from the effective model in Eq.~\eqref{eq:effective_H} to the dispersion calculated by numerically diagonalizing the Hamiltonian $H$ in Eq.~\eqref{eq:Hamiltonian}. For the numerical calculations, we project $H$ onto the lowest 20 states $\psi_{n}(r)$ in the radial direction, see Eq.~\eqref{eq:basis-states}, and consider 40 total angular momentum eigenstates. We observe an excellent agreement between effective theory and the numerics even at rather large values of momentum $p_z$, strain $\epsilon_{r,z}$ and  thickness $\tau$. 

At $p_z=0$, the eigenstates of the effective Hamiltonian $H_\text{GS}$ in Eq.~\eqref{eq:effective_H} are the degenerate Kramers partners
\begin{equation}
\label{eq:eigenstates_pz0}
|g_{1,2}^l\rangle= \frac{|\uparrow\rangle\mp|\downarrow\rangle}{\sqrt{2}}|\pm l\rangle \ , \ \text{and} \ |e_{1,2}^l\rangle=\frac{|\uparrow\rangle\pm|\downarrow\rangle}{\sqrt{2}}|\pm l\rangle \ ,
\end{equation}
 where $|\uparrow\downarrow\rangle$ are the pseudospins and $| \pm l\rangle$ are the total angular momentum eigenstates with eigenvalues $p_\varphi~|~\pm~l~\rangle=\pm\hbar (l-1/2)|~\pm~l \rangle $, with $l\in\mathbb{N}$. The eigenenergies are 
\begin{equation}
\epsilon_{g,e}= \frac{\hbar^2(2l-1)^2}{8m_\varphi R^2}\mp\frac{\hbar}{2R}\left(v_--v_+\right)\left(2l-1\right) \ ,
\end{equation}
where the angular mass is $m_\varphi=\left({1}/{m^*}-{1}/{\delta m}\right)^{-1}= m /\left[\gamma_1+\gamma_s-3(\tilde{\gamma}+\gamma_s\tilde{\epsilon}_z) \right ]$, see Eq.~\eqref{eq:pars}; neglecting the small effect of strain, we find  $m_\varphi \approx 0.06 m$, thus resulting in a variation of the angular mass of $20\%$ from the average value $m/\gamma_1$.

The ground state of the system comprises the Kramers doublet  $|g^{l=1}_{1,2}\rangle$, and is  separated from the first excited doublet $|e^{l=1}_{1,2}\rangle$ by the SOI energy
\begin{equation}
\label{eq:delta}
\Delta=\frac{\hbar }{R}v_\varphi= \frac{3}{2}\frac{\tilde{\kappa}}{\gamma_1}\hbar \omega_\varphi   \ ,
\end{equation}
where we introduce the angular SOI velocity $v_\varphi=v_--v_+$ and the dimensionless quantity
 \begin{equation}
 \label{eq:tildekappa}
\tilde{\kappa}= \gamma _s-\tilde{\gamma}- \left(\gamma _1+\gamma _s\right)\tilde{\epsilon}_z \ .
\end{equation}
The amplitude of the energy splitting is set by the characteristic angular momentum quantization energy $\hbar\omega_\varphi=\hbar^2\gamma_1/mR^2\approx 10\times (10~\text{nm}/R)^2$~meV, but it is reduced by the strain-dependent SOI $v_\varphi m R/\hbar\gamma_1=3\tilde{\kappa}/2\gamma_1$, which in parameter range considered ranges from $\sim 0.5$ at $\epsilon_z=0$, to $\sim 0.15$ at $\epsilon_z=0.5\epsilon_r$, see Fig.~\ref{fig:pars}. 
Importantly, the longitudinal strain $\epsilon_z$ \textit{decreases} the subband gap $\Delta$. This trend is in striking contrast to Ge/Si core/shell nanowires, where the strain $\epsilon_z$ induced by the Si shell \textit{increases} the energy gap at $p_z=0$, resulting in a gap $\Delta_\text{c/s}\approx 0.5 \epsilon_z\sim 10$~meV~\cite{PhysRevB.90.115419,DRkloeffel1,DRkloeffel2,DRkloeffel3}.\\

To define spin qubits, a subband energy gap $\Delta$ in the meV range is convenient because it reduces the leakage outside of the computational subspace. In strongly strained thin Ge CQWs, this requirement would constrain the maximal values of the radius $R$ to 10-20 nm.
While technologically feasible, this requirement can be also relaxed by including an external electric field $E_x$, e.g. produced by a back or top gate. This field breaks the rotational symmetry and yields the additional energy
\begin{equation}
\label{eq:E-field_H}
H_E= -eE_xR\cos(\varphi) \ ,
\end{equation}
which introduces matrix elements  with amplitude $eE_xR/2$ coupling  eigenstates of the total angular momentum quantum number $l$ to states with $l\pm 1$. 

As shown in  Fig.~\ref{fig:band}b), the effective Hamiltonian $H_\text{GS}+H_E$ reproduces nicely the energy dispersion of the low-energy states even in the presence of rather large electric fields. As anticipated, $E_x$ also induces a large energy splitting between the ground state and the first excited state at $p_z=0$.
At large values of $E_x$, the angular coordinate $\varphi$ is pinned in the vicinity of $\varphi=0$ and thus the wavefunction is confined at the top of the well. In this case, one can expand $H_E$ close to $\varphi=0$, resulting in  
\begin{equation}
\label{eq:HO_expansion}
H_\text{GS}(p_z=0)+H_E\approx \frac{p_\varphi^2}{2m_\varphi R^2}+\frac{eE_xR}{2}\varphi^2 + \Delta p_\varphi \sigma_x \ .
\end{equation}
Because the mass $m_\varphi\approx 0.06 m$ is to good approximation independent of strain, see Fig.~\ref{fig:pars}, the harmonic confinement frequency 
\begin{equation}
\label{eq:omegaE}
\omega_E=\sqrt{\frac{eE_x}{m_\varphi R}} \ 
\end{equation}
 is independent of strain and dominates over the smaller gap $\Delta$. In this case, the subband gap between the ground and first excited doublet is to good approximation $\hbar\omega_E\approx 8$~meV at $E_x=1$~V/$\mu$m and $R=20$~nm, and it decreases slowly $\propto R^{-1/2}$ compared to the faster $R^{-2}$ decay of $\Delta$.
To facilitate the comparison to Fig.~\ref{fig:band}b), we note that the ratio $\frac{\hbar\omega_E}{\epsilon_c} = \frac{\tau}{\pi R} \sqrt{\frac{eE_xR}{\epsilon_c}} \sqrt{\frac{m}{\gamma_1 m_\varphi}}\approx 0.17$ at $\tau=R/2$ and $eE_xR=\epsilon_c$, in good agreement with the figure.

We remark that the electric field $E_x$ originates from a gate that breaks rotational symmetry and it cannot come from gates completely wrapped around the CQW. Such gates all-around will instead produce a radial electric field $E_r$ that does not break the inversion symmetry, but that can quantitatively modify the effective parameters in Eq.~\eqref{eq:pars}. We also note by a third order Schrieffer-Wolff transformation that  this electric field  generates a \textit{cubic} SOI term
\begin{equation}
H_{r}= \frac{-4 \tilde{\gamma} \gamma _s eE_r\tau}{\pi ^4 \left(\gamma _1-2 \gamma _s\right) \left( 2\gamma _s+\gamma _1 \left(2 \epsilon _r+\epsilon _z\right)/\epsilon _c\right)}\frac{p_-^3\tau^3}{\hbar^3}\sigma_+ +\text{h.c.}  \ ,
\end{equation}
in analogy to planar Ge/SiGe heterostructures~\cite{PhysRevB.103.125201,Wang2021,bosco2021squeezed}.
While this term can be of interest at large values of $R$, in the regime of parameters studied in this work it only adds a small correction and will not be discussed further.\\

Finally, we introduce the magnetic field $\textbf{B}$ in the effective theory.
By writing the Hamiltonian $H_\textbf{B}$ in the radial basis in Eq.~\eqref{eq:basis-states}, and with a second order Schrieffer-Wolff transformation, we find to linear order in $\textbf{B}$
\begin{equation}
\label{eq:Zeeman}
\begin{split}
\frac{H_B}{3\kappa\mu_B}&=  B_x \left\{\cos(\varphi)\left[\left(1-\frac{\tilde{\gamma}}{\kappa}\right)\sigma_z-\frac{\tilde{\kappa}}{\kappa}\frac{z}{R} \sigma_x \right]+\tilde{\epsilon}_z\sin(\varphi)\sigma_y\right\}\\
&+B_y \left\{\sin(\varphi)\left[\left(1-\frac{\tilde{\gamma}}{\kappa}\right)\sigma_z-\frac{\tilde{\kappa}}{\kappa}\frac{z}{R} \sigma_x \right]-\tilde{\epsilon}_z\cos(\varphi)\sigma_y\right\}\\ 
&+ B_z \left[\left(\tilde{\epsilon}_z+\frac{1}{2}\frac{\tilde{\kappa}}{\kappa}\right)\sigma_x+\frac{1}{3\kappa} \frac{m}{m_\varphi}p_\varphi   \right] \ .
\end{split}
\end{equation}
We omit  the negligible spin-independent shift of the  dot $-2\mu_B z\left\{p_\varphi,[B_x\cos(\varphi)+B_y\sin(\varphi)]\right\}/m_\varphi R$.
We emphasize that  the magnetic interactions have an angular dependence caused by the transformation $U$ and that the origin of the coordinate system coincides with the center of mass of the quantum dot. In addition, the corrections to the Zeeman energy caused by the high energy  holes are discussed in detail in App.~\ref{app:SOHs}, see in particular Eq.~\eqref{eq:Zeeman_SOH}.

\section{Spin qubits in Short quantum dots}

\label{sec:SQD}
We now study the properties of a spin qubit confined in a quantum dot in the thin CQWs sketched in Fig.~\ref{fig:sketch}.
The behaviour of the spin qubit strongly depends on the length $l_z$ of the  dot. 
In particular, we examine two different qubit designs where the  dot is long and short compared to the radius $R$, i.e. $l_z\gg R$ and $l_z\lesssim R$, respectively. Both these regimes can be described by the  effective theory introduced in Sec.~\ref{sec:Effective-theory}.
At first, we restrict ourselves to the analysis of the annular CQW shown in Fig.~\ref{fig:sketch}a) and described  by the isotropic LK Hamiltonian in Eq.~\eqref{eq:LK-Hamiltonian}. We then show that our theory well describes also the planar CQW in Fig.~\ref{fig:sketch}b). Moreover, our theory models a wide range of devices with general cross-sections and is valid even when cubic anisotropies of the LK Hamiltonian are included. A detailed analysis of  anisotropic corrections, including results obtained for quantum wells grown along a main crystallographic axis and with square cross-sections, is given in App.~\ref{sec:deviation_LK}.

The dynamics of a spin qubit can be mapped to the effective quantum dot Hamiltonian~\cite{doi:10.1063/1.4858959,PhysRevLett.120.137702,PhysRevApplied.16.054034,VenitucciElectricalmanipulationsemiconductor2018,doi:10.1126/science.1080880}
\begin{equation}
\label{eq:QD-effective-theory}
H_\text{QD}=\frac{\mu_B}{2}  \pmb{\sigma}\cdot\left(\underline{g}-\sum_{i=x,y,z}\frac{eE_i(t)R}{\hbar\omega_\varphi}\delta\underline{ g}^i\right)\cdot\textbf{B} \ ,
\end{equation}
parametrized by a tensor $\underline{g}$ of $g$-factors  and an tensor $ \delta\underline{g}^i$  driving spin transitions via the ac field $E_i(t)$. 

An accurate model to describe short quantum dots, with $l_z\lesssim R$, is provided by the Hamiltonian
\begin{equation}
\label{eq:SQD_theory}
H_\text{SD}=\frac{\hbar\bar{\omega}_\varphi}{2}p_\varphi^2+\Delta p_\varphi\sigma_x -\frac{\hbar\bar{\omega}_z}{2} \lambda_z+\frac{\hbar v_z}{\sqrt{2}\bar{l}_z}  \lambda_y\sigma_y-eE_xR\cos(\varphi) \ ,
\end{equation}
obtained by projecting Eqs.~\eqref{eq:effective_H} and~\eqref{eq:E-field_H} onto the first two eigenstates of the longitudinal confinement, energetically separated by  $\hbar\bar{\omega}_z\propto 1/\bar{l}_z^2$.
 We introduce the longitudinal mass $m_z=\left({1}/{m^*}+{1}/{\delta m}\right)^{-1}\approx 0.06m$ and SOI velocity $v_z=v_-+v_+\approx 6.5 \hbar/mR$, see Eq.~\eqref{eq:pars} and Fig.~\ref{fig:pars}. Strain weakly affects these parameters and in the regimes studied it results in variations of $\lesssim 15\%$ from the numerical values provided here.
We also introduce the exact longitudinal and angular frequencies $ \bar{\omega}_z=\omega_z\sqrt{m/\gamma_1 m_z} $ and $ \bar{\omega}_\varphi=\omega_\varphi\sqrt{m/\gamma_1 m_\varphi}$, that include the small  corrections of longitudinal and angular masses from the average value $m/\gamma_1$; in analogy, we define the exact harmonic length $\bar{l}_z=l_z(\gamma_1 m_z/m)^{1/4}$.
 Here, the Pauli matrices $\lambda_{x,y,z}$  act on these orbital states, while   $\sigma_{x,y,z}$ act on the pseudo spin. We remark that $p_\varphi$ is the total angular momentum, and that at $E_x=0$ the degenerate Kramers partners are the eigenstates of $p_\varphi \sigma_x$ (and not of $\sigma_x$) to the same eigenvalue, see Eq.~\eqref{eq:eigenstates_pz0}.

To proceed further, it is  convenient to eliminate the term  $\propto\lambda_y\sigma_y$ by the rotation $H_\text{SD}\to e^{i \theta_z \lambda_x\sigma_y/2} H_\text{SD}e^{-i \theta_z \lambda_x\sigma_y/2} $ where $\theta_z=\arctan(\sqrt{2}v_z/\bar{\omega}_z\bar{l}_z)$. After this transformation, the low energy Hamiltonian acting on the ground state Kramers partners is 
\begin{equation}
\label{eq:gs-SQD_theory}
H_\text{SD}^\text{GS}=\frac{\hbar\bar{\omega}_\varphi}{2}p_\varphi^2+\Delta\cos(\theta_z) p_\varphi\sigma_x -eE_xR\cos(\varphi) \ .
\end{equation} 
At weak electric fields, the subband gap that energetically separates the spin qubit to the non-computational subspace is
\begin{equation}
\label{eq:Eg_SQD}
E_g=\sqrt{\Delta^2 \cos(\theta_z)^2+e^2E_x^2R^2} \ .
\end{equation}
In quantum wells with $R=20$~nm, the gap is $E_g\approx 1$~meV at $E_x=0$ and it increases with $E_x$. 
 At large values of $E_x\gtrsim 1$~V/$\mu$m, the energy gap approaches $\hbar \omega_E\gtrsim 8$~meV, see Eq.~\eqref{eq:omegaE}.
Also, in this basis, Eq.~\eqref{eq:Zeeman} reduces to
\begin{equation}
\label{eq:Zeeman_SQD}
\begin{split}
H_\text{SD}^{B}=& 3\kappa\mu_B B_z \left[\left(\tilde{\epsilon}_z+\frac{1}{2}\frac{\tilde{\kappa}}{\kappa}\right)\cos(\theta_z)\sigma_x+\frac{1}{3\kappa} \frac{m}{m_\varphi}p_\varphi   \right]\\
&+ 3\kappa\mu_B\left[B_x\cos(\varphi)+B_y\sin(\varphi) \right]  q_0  \sigma_z\\
&+3\kappa\mu_B\left[B_x\sin(\varphi)-B_y\cos(\varphi) \right]\tilde{\epsilon}_z\sigma_y \ ,
\end{split}
\end{equation}
where we introduce the size-dependent quantity
\begin{equation}
q_0=\left(1-\frac{\tilde{\gamma}}{\kappa}\right)\cos(\theta_z)-\frac{\tilde{\kappa}}{\kappa}\frac{\bar{l}_z}{\sqrt{2}R}\sin(\theta_z) \ .
\end{equation}

\begin{figure}
\centering
\includegraphics[width=0.45\textwidth]{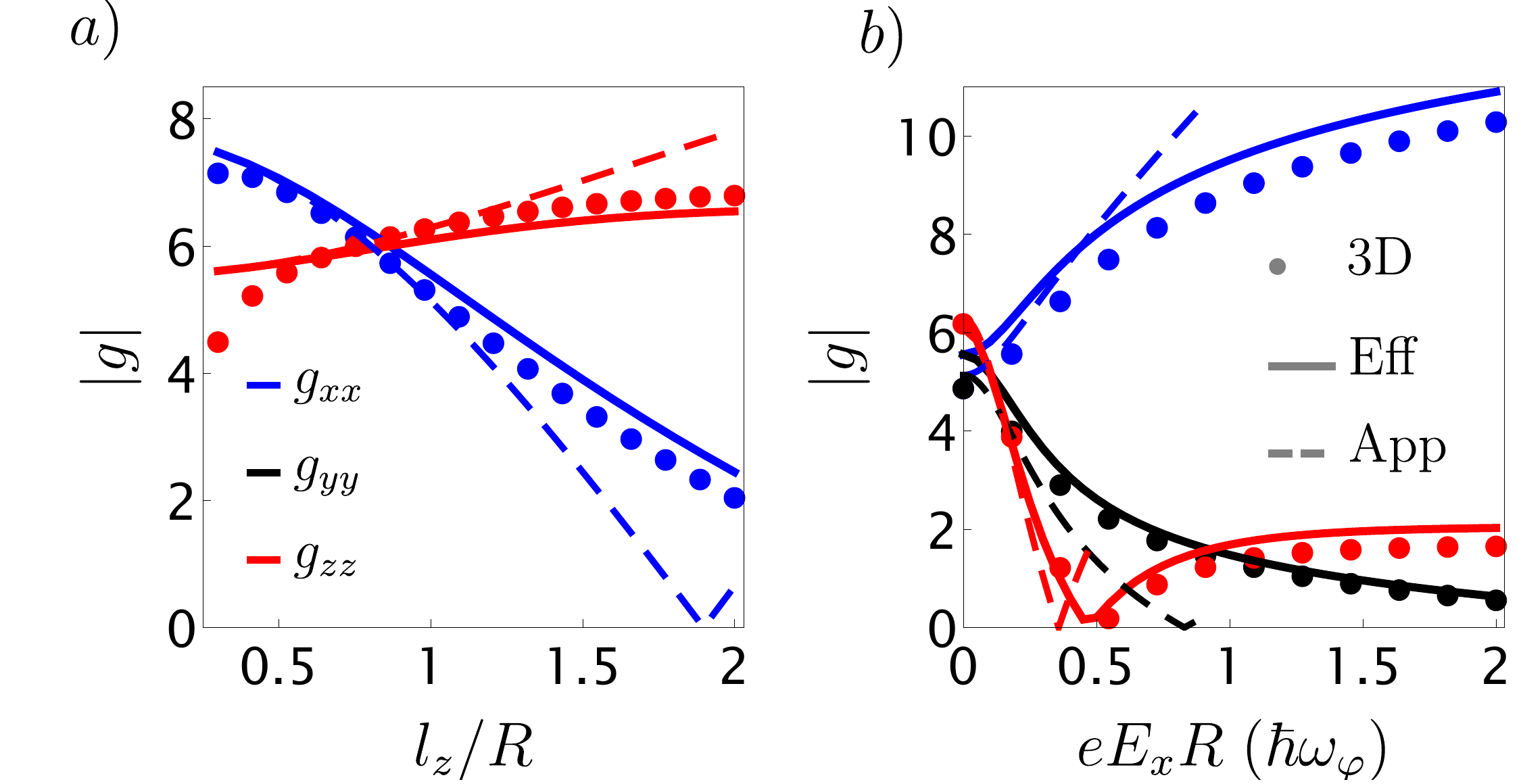}
\caption{\label{fig:gfact_SQD}
Matrix of $g$-factors of a hole spin qubit in a thin Ge CQW. We show the diagonal elements $g_{xx}$, $g_{yy}$, $g_{zz}$ in blue, black and red, respectively; the off-diagonal elements are zero.
We compare the results of a numerically simulation a three-dimensional quantum dot obtained by discretizing the Hamiltonian~\eqref{eq:Hamiltonian} with an annular cross section (dots), against the effective theory~\eqref{eq:SQD_theory} (solid lines) and the approximate formulas in Eq.~\eqref{eq:g-tens_app} (dashed lines).
In a), we show $g_{ii}$ as a function of the length $l_z$ of the quantum dot at $E_x=0$; in this case, $|g_{xx}|=|g_{yy}|$.
In b), we show $g_{ii}$ as a function of the electric field $E_x$ at $l_z=R$. In both cases, we consider $\epsilon_r=3\epsilon_z=\epsilon_c$  and $\tau=R/2$.
At $R=20$~nm, the electric field is $E_x\in[0,0.26]$~V/$\mu$m. }
\end{figure}

\subsection{Defining the spin qubit }

We now discuss the matrix $\underline{g}$ of $g$-factors, that determines the energy gap of different spin states in the quantum dot.
From Eqs.~\eqref{eq:gs-SQD_theory} and~\eqref{eq:Zeeman_SQD},  $\underline{g}$ can be derived by projecting $H_\text{SD}^{B}$ onto the degenerate groundstate of $H_\text{SD}^\text{GS}$.
The resulting $g$-tensor is diagonal and $\underline{g}=g_{ii}\delta_{ij}$. The diagonal element $g_{ii}$ are shown in Fig.~\ref{fig:gfact_SQD}.
We note that the effective theory in Eq.~\eqref{eq:SQD_theory} is reasonably accurate and approximates well a numerical simulation of a three-dimensional quantum dot obtained from Eq.~\eqref{eq:Hamiltonian}.
In Fig.~\ref{fig:gfact_SQD}a), we shown the dependence of $g_{ii}$ on the size of the dot at $E_x=0$. In this case, the eigenstates of Eq.~\eqref{eq:gs-SQD_theory} coincide with the eigenstates of $p_\varphi \sigma_x$ given in Eq.~\eqref{eq:eigenstates_pz0} and one obtains $g_{xx}=-g_{yy}= -3\kappa (q_0-\tilde{\epsilon}_z)$ and $g_{zz}= 6\kappa(\tilde{\epsilon}_z +{\tilde{\kappa}}/2\kappa)\cos(\theta_z)-{m}/{m_\varphi}$.

Because of the orbital magnetic field,  $g_{zz}$ is rather large, and only weakly dependent on the length $l_z$ of the dot.  A similar enhancement of effective Zeeman energy emerges in topological insulator nanowires~\cite{PhysRevB.104.165405}, where the leading contribution is the SOI-induced term $\propto \tilde{\kappa}$. Because in competing hole-based architectures, such as Ge/Si core/shell nanowires, $g_{zz}\sim 1$ is rather small~\cite{adelsberger2}, the  large value of $g_{zz}$ in thin CQWs is particularly advantageous for topological quantum computing in the seek of exotic particles, such as Majorana bound states~\cite{PhysRevB.86.085408,PhysRevB.90.195421}.

At small values of $E_x$, one can still use a few eigenstates in Eq.~\eqref{eq:eigenstates_pz0}. By using the energetically lowest 2 quasi-degenerate Kramers partners and by introducing the $E_x$-dependent angle $\theta_E=\arctan\left[eE_xR/\Delta\cos(\theta_z)\right]$, one finds to second order in perturbation theory 
\begin{subequations}
\label{eq:g-tens_app}
\begin{align}
g_{xx}&=3\kappa\left[
\tilde{\epsilon}_z \cos(\theta_E)-q_0-q_0\frac{m_\varphi}{m}eE_xR \sin(\theta_E)\right]\\
g_{yy}&=3\kappa\left[
q_0 \cos(\theta_E)-\tilde{\epsilon}_z-\tilde{\epsilon}_z\frac{m_\varphi}{m}eE_xR \sin(\theta_E)\right] \\
g_{zz}&=6\kappa\left(\tilde{\epsilon}_z +\frac{\tilde{\kappa}}{2\kappa}\right)\cos(\theta_z)-\frac{m}{m_\varphi}\cos(\theta_E) \ .
\end{align}
\end{subequations}
These equations capture qualitatively the trend of the $g$-factors as a function of $E_x$, but they are not quantitatively accurate at large values of $E_x$, as shown in Fig.~\ref{fig:gfact_SQD}b).

At large values of $E_x$, Eq.~\eqref{eq:gs-SQD_theory} is approximated by a harmonic oscillator with harmonic frequency $\omega_E$, see Eqs.~\eqref{eq:HO_expansion} and~\eqref{eq:omegaE}, resulting in
$g_{xx}=-6\kappa e^{-\varphi_E^2(\varphi_{S}^{-2}+\frac{1}{4})}[q_0\cosh(\varphi_E^2/\varphi_{S})-\tilde{\epsilon}_z\sinh(\varphi_E^2/\varphi_{S})]$, $g_{yy}=6\kappa e^{-\varphi_E^2(\varphi_{S}^{-2}+\frac{1}{4})}[q_0\sinh(\varphi_E^2/\varphi_{S})-\tilde{\epsilon}_z\cosh(\varphi_E^2/\varphi_{S})]$ and
$g_{zz}=6\kappa\tilde{\epsilon}_z\cos(\theta_z)$ with angular width 
$\varphi_E=\sqrt{\omega _{\varphi } m/ \gamma _1  m_{\varphi }\omega _E}$ and SOI angle $\varphi_{S}^{-1}=m_\varphi v_\varphi\cos(\theta_z)R/\hbar$~\footnote{The signs of the $g$-factors can be understood from Eq.~\eqref{eq:Zeeman_SQD} by considering a rotation of $\pi/2$ around $y$, that transforms $\sigma_x\to \sigma_z$ and $\sigma_z\to -\sigma_x$. In addition, at large $E_x$, because of the SOI $\Delta \cos(\theta_z)\sigma_x p_\varphi $ in the effective Hamiltonian~\eqref{eq:SQD_theory}, the expectation value of $m p_\varphi/m_\varphi$ in the harmonic oscillator groundstate exactly cancels the SOI-induced Zeeman energy $\propto \tilde{\kappa}$  in the estimation of $g_{zz}$. One also finds that $\cos(\varphi)\sigma_{y,z}\to \pm e^{-\varphi_E^2(\varphi_{S}^{-2}+\frac{1}{4})}\cosh(\varphi_E^2/\varphi_{S})\sigma_{y,x} $ and $\sin(\varphi)\sigma_{y,z}\to  e^{-\varphi_E^2(\varphi_{S}^{-2}+\frac{1}{4})}\sinh(\varphi_E^2/\varphi_{S})\sigma_{x,y} $.}.
Because, depending on the amplitude of the longitudinal strain $\epsilon_z$, the limiting values of the $g$-factor for $E_x=0$ and $E_x\to \infty$ can have opposite signs, there could be points at finite values of $E_x$ where the $g_{ii}$ vanish, as shown in Fig.~\ref{fig:gfact_SQD}b).

We note that the $g$-tensor is strongly anisotropic and  it is tunable by the electric field and by designing the strain, especially via the outer shell thickness. This behaviour is typical of hole nanostructures~\cite{PhysRevLett.110.046602,PhysRevLett.127.190501,adelsberger2021hole,doi:10.1063/1.5025413,PhysRevB.87.161305,qvist2021anisotropic}. In contrast to Ge/Si core/shell nanowires~\cite{adelsberger2021hole,froning2020ultrafast}, however, in a thin CQW the $g$-factor is only strongly modulated at weak values of the electric field $E_x$, and at $E_x\gtrsim 1$~V/$\mu$m, $g$ becomes  weakly dependent of $E_x$. In particular, $g$ is \textit{independent} of $E_x$ when $\textbf{B}$ is aligned to the quantum well and it only varies as $E_x^{-1/2}$ when $\textbf{B}$ is perpendicular to it. This property suggests that spin qubits in these devices can have a low susceptibility to charge noise, enabling long coherence times.

\subsection{Decoherence of the qubit}

\label{sec:decoh}
\begin{figure}
\centering
\includegraphics[width=0.5\textwidth]{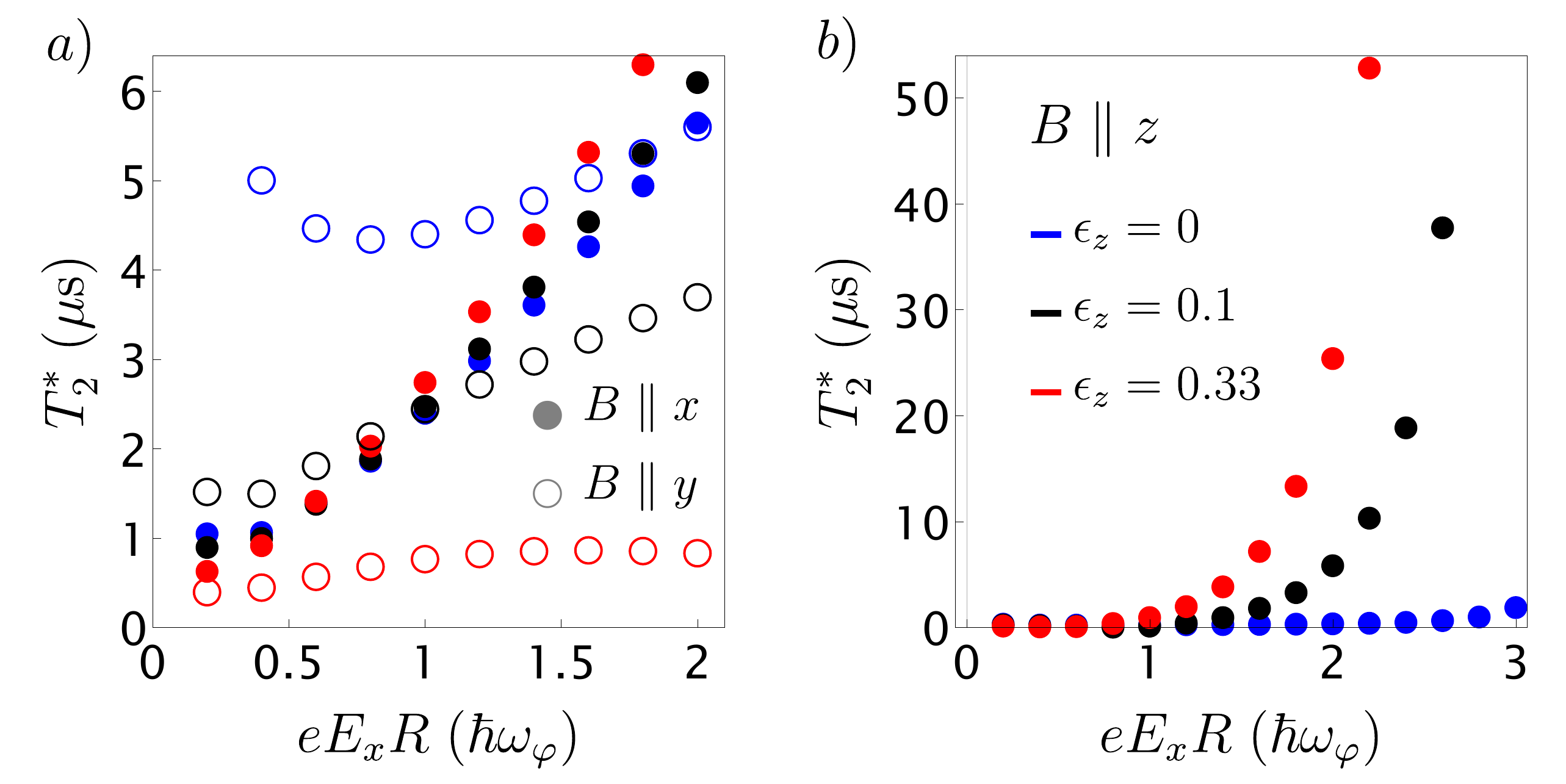}
\caption{\label{fig:deph} Dephasing time $T_2^*$ of a hole spin qubit in a thin CQW as function of electric field $E_x$. In a), we show with full and empty dots the time $T_2^*$ when the magnetic field is aligned to the $x$ and $y$-directions, respectively; in b), we show the result when $\textbf{B}\parallel z$. Blue, black and red dots indicate longitudinal strain $\epsilon_z=0$, $\epsilon_z=\epsilon_c/10$, and $\epsilon_z=\epsilon_c/3$, respectively. For the simulation, we consider a free induction decay experiment, in a qubit with frequency $\omega_Q^i/2\pi=5$~GHz, and  $\alpha \sqrt{\delta V^2}=0.3$~$\mu$eV. We also use $R=l_z=\tau/2=20$~nm, and $\epsilon_r=\epsilon_c$.
 }
\end{figure}

We now discuss the coherence time of these hole spin qubits. We examine a free induction decay experiment, where the spin is prepared in an eigenstate of the Zeeman energy and left idle.
In this case, 1/f charge noise causes random fluctuations of the electric potential, with  spectral function $S(\omega)=\langle\delta V^2\rangle/|\omega|$.
We consider here magnetic fields aligned to the main confinement axes $i=(x,y,z)$.
Because of the dependence of $g$ on the external electric field, the charge noise causes dephasing, with decay rate~\cite{MAKHLIN2004315,bosco2021squeezed,bosco2020hole}
\begin{equation}
\label{eq:t2star}
\frac{1}{T_2^*}\approx \frac{\omega_Q^i}{\sqrt{2\pi}} \frac{1}{g_{ii}} \frac{\partial g_{ii}}{\partial V}  \sqrt{ \langle \delta V^2\rangle} \ .
\end{equation}  
We neglect small logarithmic corrections caused by the divergence of the spectral function at low frequency~\cite{MAKHLIN2004315}.
The qubit frequency $\omega_Q^i=g_{ii}\mu_B B_i/\hbar$ depends on the external magnetic and electric fields and we restrict our analysis to microwave frequencies $\omega_Q^i/2\pi\sim 1-20$~GHz. 
To estimate the sensitivity of the $g$-factor to the potential fluctuations, we assume that the noise comes from the electrodes, such that ${\partial g_{ii}}/{\partial V} \approx  \alpha\delta g_{ii}^x /\hbar\omega_\varphi$, see Eq.~\eqref{eq:QD-effective-theory}. Here,  $\alpha\sim 0.1-0.5$ is the lever arm of the gate~\footnote{If $\mu$ is the chemical potential of the dot, $\partial_V g_{ii} = \alpha \partial_\mu g_{ii}\approx  \alpha \delta^x_{ii}g_{ii}/\hbar\omega_\varphi $. We use here  the  definition of the lever arm $\alpha=\partial\mu/\partial V$~\cite{burkard2021semiconductor},  of $\delta g^x_{ii}=\hbar\omega_\varphi\partial_{E_x} g_{ii}/eR$, and we assume that the variation of the chemical potential are caused only by $E_x$, i.e. $\Delta\mu\approx eE_xR$.}, whose precise value depends on the device design; resulting in the typical values  $\alpha \sqrt{\langle\delta V^2\rangle}\sim 0.1-10$~$\mu$eV~\cite{Yonedaquantumdotspinqubit2018,burkard2021semiconductor}.

In Fig.~\ref{fig:deph}, we show the decay time $T_2^*$ of a typical hole spin qubit in a thin CQW as a function of the applied electric field. 
When the magnetic field is applied in the $(x,y)$ plane, $T_2^*$ is generally in the $\mu$s range, and its precise value depends on $E_x$ and on the strain $\epsilon_z$. In particular, when $\textbf{B}\parallel E_x$ the strain dependence  of $T_2^*$ is to good approximation negligible  and $T_2^*$ increases monotonically with $E_x$. In contrast, when $\textbf{B}\parallel y$, the electric field dependence is weaker, and $T_2^*$ is strongly affected by $\epsilon_z$. The decrease of $T_2^*$ as strain increases  originates from the reduced value of $g_{yy}$ at large $\epsilon_z$ and large $E_x$, see Eq.~\eqref{eq:t2star}. For example at $\epsilon_z=0.33\epsilon_c$ and  $E_x= \hbar\omega_\varphi/eR\approx 0.13$~V/$\mu$m at $R=20$~nm, one finds $T_2^*\approx 750$~ns for the parameters considered. This dephasing time can be further improved by echo sequences~\cite{PhysRevLett.100.236802,Bluhm2011}.

Strikingly, the flatness of the  $g$-factor in the $z$-direction as a function of $E_x$ results in rather long coherence time of qubits in strained CQWs, with an enhanced value of $T_2^*$ when $E_x\gtrsim 2 \hbar\omega_\varphi/eR\approx 0.26$~V/$\mu$m at $R=20$~nm. At low values of $\epsilon_z$, the small value of $g_{zz}\propto\tilde{\epsilon}_z$ at large $E_x$ reduces $T_2^*$.  Because $g_{zz}$ is only weakly dependent on the quantum dot length $l_z$, see Fig.~\ref{fig:gfact_SQD}a), this enhancement occurs also in long quantum dots.
We emphasize that, in contrast to alternative proposals where the lifetime of the qubit is enhanced only at fine-tuned sweet-spots~\cite{bosco2020hole, Wang2021,Malcok2022}, in thin Ge CQWs the qubit is to good approximation insensitive to charge noise in a wide range of $E_x$, thus enabling highly coherent qubits with a low sensitivity to charge noise,  a major issue in state-of-the-art hole spin quantum processors~\cite{froning2020ultrafast}.

\subsection{Driving the qubit}

\label{sec:driving}
\begin{figure}
\centering
\includegraphics[width=0.5\textwidth]{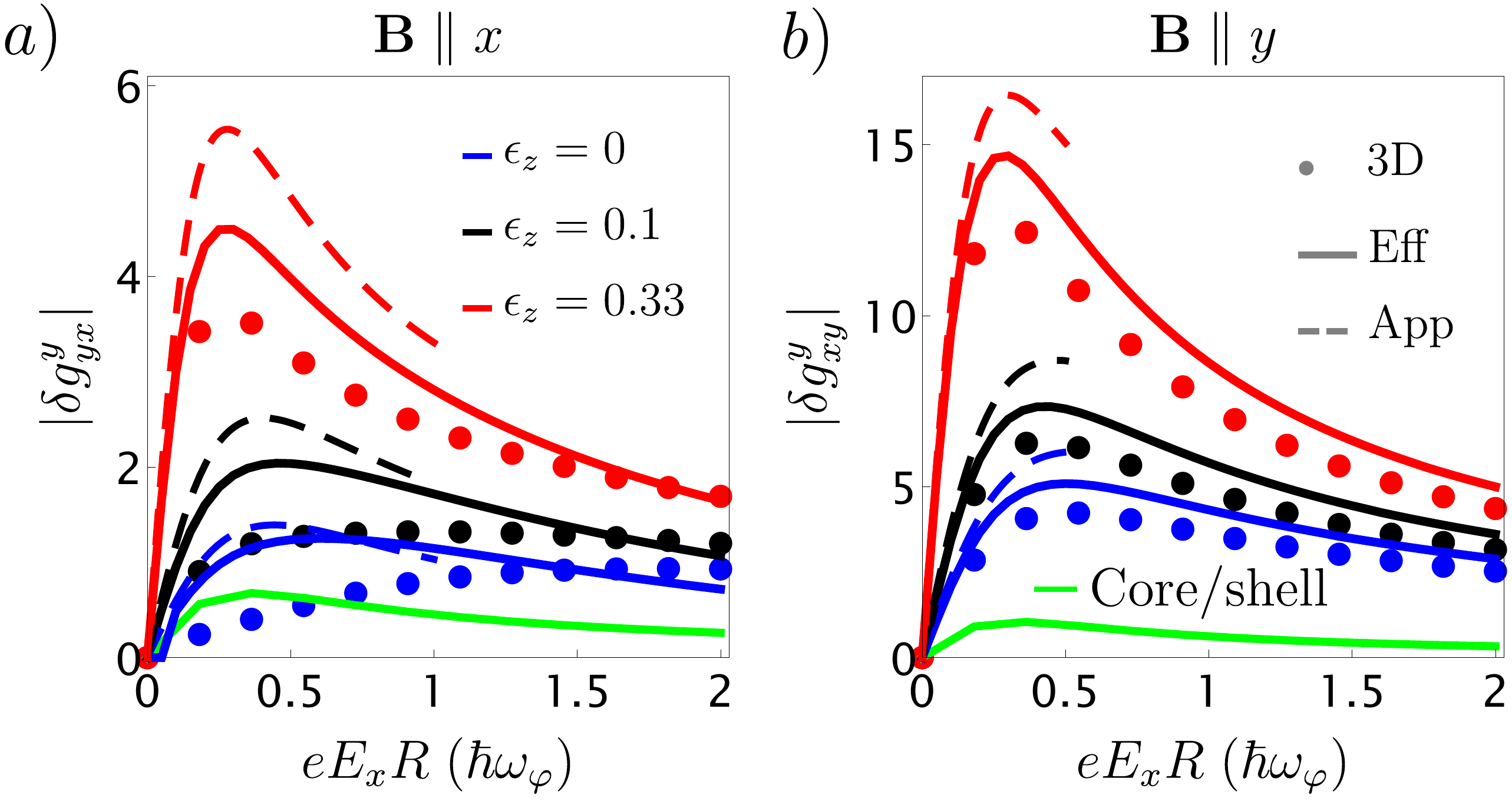}
\caption{\label{fig:drive_SQD}
Driving of a hole spin qubit in thin Ge CQW by an ac electric field $E_y(t)$. In a) and b), we show the non-zero elements $\delta g_{yx}^y$ and $\delta g_{xy}^y$ of the matrix $\delta \underline{g}^y$ as a function of the dc electric field $E_x$, see Eq.~\eqref{eq:QD-effective-theory}. With blue, black, and red curves, we show different values of longitudinal strain $\epsilon_z=0$, $\epsilon_z=\epsilon_r/10$, and $\epsilon_z=\epsilon_r/3$, respectively.   We use $\tau=R/2$, $l_z=R$ and $\epsilon_r=\epsilon_c$. The dots, the solid and the dashed lines represent results obtained by Eqs.~\eqref{eq:Hamiltonian},~\eqref{eq:SQD_theory}, and~\eqref{eq:delta_g_app}, respectively. For the solid and dashed lines, we use the two lowest Kramers partners at $E_x=0$. The green lines show the results obtained by simulating a Ge/Si core/shell nanowire with $\epsilon_z=\pi^2\hbar\omega_\varphi/4$, where this effect is negligible. }
\end{figure}

We now discuss two qualitatively different mechanisms to drive the hole spin qubit.
First, we note that in short quantum dots, where $l_z\lesssim R$, shaking the hole wavefunction along the quantum well~\cite{PhysRevB.74.165319} does not result in ultrafast Rabi oscillations. As discussed in  Sec.~\ref{sec:Elongated-QDs}, this mechanism  is more convenient in long quantum dots because the Rabi frequency $\omega_R\propto l_z^4$~\cite{bosco2021squeezed}, see Eq.~\eqref{eq:Rabi_EQD}. However, as shown in Fig.~\ref{fig:band}b), in a CQW the electric field $E_x$ confines the hole to the top of the cross-section and thus fast Rabi oscillations are enabled when the hole is periodically driven in the angular direction by an ac  field $E_y(t)=E_y^{ac}\sin(\omega_D t)$, perpendicular to the dc field $E_x$. 

These spin transitions are parametrized by the matrix $\delta\underline{g}^y$ of effective $g$-tensors, see Eq.~\eqref{eq:QD-effective-theory}, and are the fastest when the dc field $E_x$ is comparable to the SOI gap $\Delta$; for this reason, we restrict ourselves to the analysis of the moderately weak electric fields. By including the small potential $-eE_y(t)R\sin(\varphi)$ in Eq.~\eqref{eq:gs-SQD_theory} and by using second order perturbation theory, we find that the driving term is $\propto E_y(t)(\sigma_x\delta g_{xy}^yB_y +\sigma_y\delta g_{yx}^yB_x)$, with
\begin{subequations}
\label{eq:delta_g_app}
\begin{align}
\delta g_{xy}^y&=3\kappa \sin(\theta_E) \left[ \frac{\hbar\omega_\varphi}{E_g}q_0+ \frac{m_\varphi\gamma_1}{m}\Big(q_0+\tilde{\epsilon}_z \cos(\theta_E)\Big)\right] \ , \\
\delta g_{yx}^y&=3\kappa \sin(\theta_E) \left[ \frac{\hbar\omega_\varphi}{E_g}\tilde{\epsilon}_z+ \frac{m_\varphi\gamma_1}{m}\Big(\tilde{\epsilon}_z +q_0\cos(\theta_E)\Big)\right]  \ .
\end{align}
\end{subequations}

We show these terms in Fig.~\ref{fig:drive_SQD}, and in particular we highlight their dependence on  $E_x$ and on the longitudinal strain $\epsilon_z$, that can be designed by the thickness of the outer Si shell, see Eq.~\eqref{eq_strain_pars}. We note that at small values of $E_x$, Eq.~\eqref{eq:delta_g_app} approximates well the driving $\delta g_{xy}^y$ when $\textbf{B}\parallel y$, but are quantitatively inaccurate for $\delta g_{yx}^y$ when $\textbf{B}\parallel x$, especially at small $\epsilon_z$.  The latter term is generally smaller and thus to enhance the Rabi frequency it is convenient to align $\textbf{B}$ to the ac drive.  Moreover, $\epsilon_z$ strongly speeds up the driving, in sharp contrast to elongated dots in Ge/Si core/shell nanowires, where the Rabi frequency at small electric fields is $\propto 1/\epsilon_z$~\cite{DRkloeffel1,DRkloeffel2,DRkloeffel3}.
We also note that while in principle a similar driving mechanism can also occur in core/shell nanowires, in the parameter range considered, the terms $\delta g_{ij}^y$ are negligible there. When $\textbf{B}\parallel z$, this effect vanishes.

We now estimate the frequency $\omega_R$ of the Rabi oscillations generated by these driving terms when $\textbf{B}$ is aligned to the $i=\{x, y\}$ directions. In this case, the spin states are split by $\hbar \omega_Q^i=g_{ii} \mu_B B_i$, and when the qubit is in resonance with the drive, i.e. $\omega_D=\omega_Q^i$, we find
\begin{equation}
\label{eq:Rabi_new-SQD}
\frac{\omega_R^i}{2\pi}=\frac{\omega_D}{2\pi} \frac{eE_y^{ac}R}{2\hbar\omega_\varphi}\frac{\delta g_{j i}^y}{g_{ii}}\approx 0.2~\text{GHz}\times\frac{\delta g_{j i}^y}{g_{ii}} \ .
\end{equation}
The numerical prefactor is obtained by considering a CQW with radius $R=20$~nm, such that $\hbar\omega_\varphi=2.5$~meV, and for the typical experimental values $\omega_D/2\pi= 5$~GHz and $E_y^{ac}=10$~mV/$\mu$m~\cite{froning2020ultrafast}. For thicker quantum wells, this factor increases as $R^3$, but the energy gap at $E_x=0$ also decreases as $1/R^2$.
By comparing Figs.~\ref{fig:gfact_SQD} and~\ref{fig:drive_SQD}, we also observe that  ${\delta g_{y x}^y}\lesssim g_{xx}$ when $\textbf{B}\parallel x$, and ${\delta g_{ xy}^y}\gtrsim g_{yy}$ when $\textbf{B}\parallel y$. For example, in a strongly strained device with a thick outer shell, at $E_x= \hbar\omega_\varphi/eR\approx 0.13$~V/$\mu$m and $R=20$~nm, one obtains $\delta g_{xy}^y/g_{yy}\approx 6$, resulting in $\omega_R^y/2\pi\approx 1.2$~GHz. At this value of $E_x$, we also find $g_{yy}=1.35$, such that $\omega_B^y/2\pi=5$~GHz at $B=0.25$~T, and the subband energy gap is $E_g=2.3$~meV. At this working point, we also expect a dephasing time of a few hundreds of nanoseconds, see Fig.~\ref{fig:deph}, much longer than the spin-flipping time, thus enabling highly coherent and ultrafast qubit operations at low power. \\

In addition, the $g$-factor can also be modulated by an ac field $E_x(t)=E_x^{ac}\cos(\omega_D t)$ applied in the $x$-direction, resulting in the variation $\delta \underline{g}^x$ of the $g$-factors in Eq.~\eqref{eq:g-tens_app}
\begin{subequations}
\label{eq:delta_g_app_Ex}
\begin{align}
\frac{\delta g_{xx}^x}{3\kappa \sin(\theta_E)}&= \frac{\hbar\omega_\varphi}{E_g}\tilde{\epsilon}_z\cos(\theta_E)+ \frac{m_\varphi\gamma_1}{2m}\Big(3+ \cos(2\theta_E)\Big)q_0\ , \\
\frac{\delta g_{yy}^x}{3\kappa \sin(\theta_E)}&= \frac{\hbar\omega_\varphi}{E_g}q_0\cos(\theta_E)+ \frac{m_\varphi\gamma_1}{2m}\Big(3+ \cos(2\theta_E)\Big)\tilde{\epsilon}_z \ , \\
\delta g_{zz}^x&=\frac{m}{2m_\varphi}\frac{\hbar\omega_\varphi}{E_g}\sin(2\theta_E)\ .
\end{align}
\end{subequations}

\begin{figure}
\centering
\includegraphics[width=0.5\textwidth]{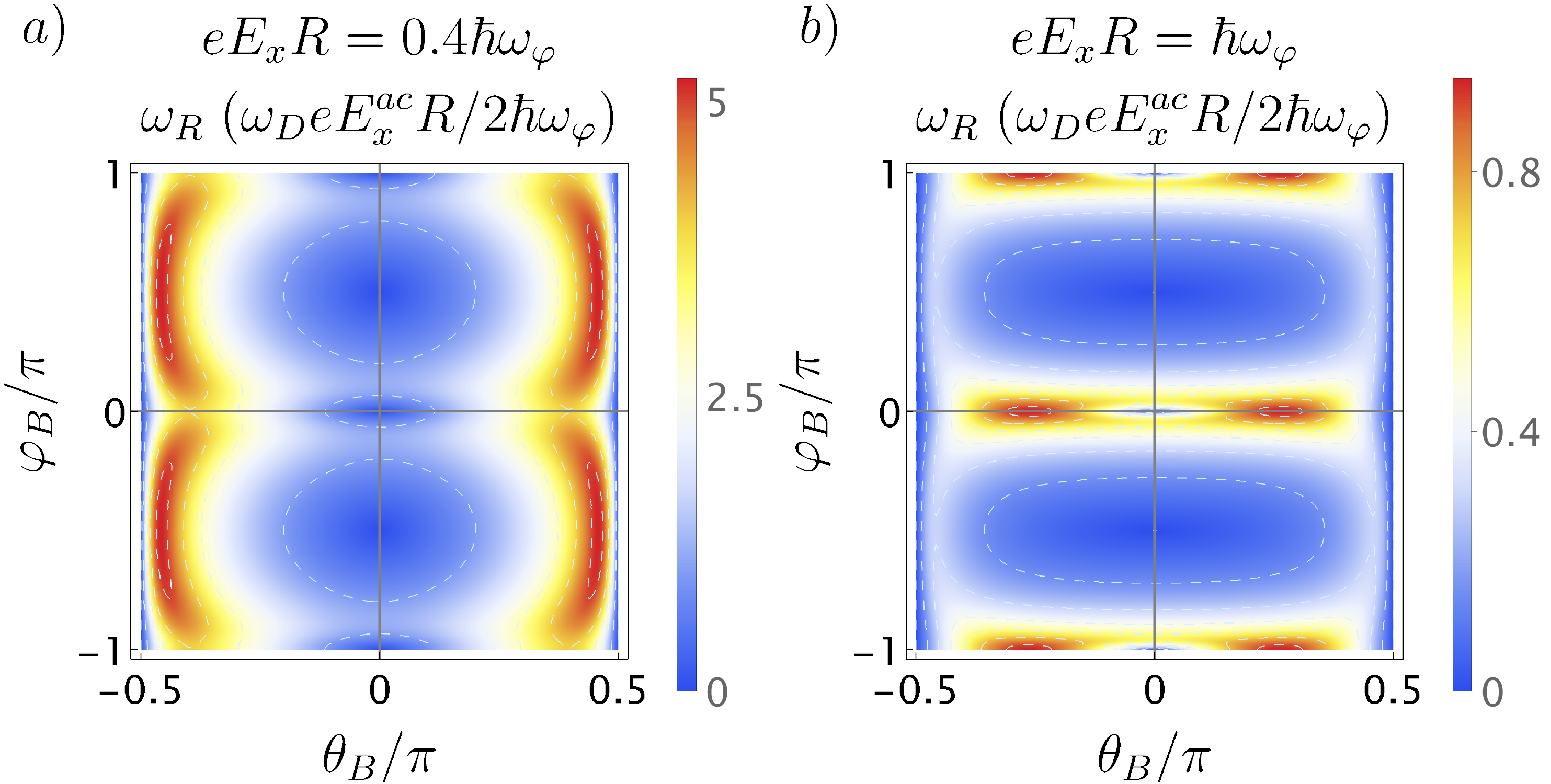}
\caption{\label{fig:drive_SQD_gmod}
Driving of a hole spin qubit in a thin Ge CQW by $g$-tensor modulation via an ac electric field $E_x^{ac}$. In a) and b), we show the Rabi frequency as a function of the direction of the applied magnetic field for two different values of the dc field $E_x$. The results shown here use Eq.~\eqref{eq_rabi-gmod} with parameters obtained by discretizing the Hamiltonian Eq.~\eqref{eq:Hamiltonian}. The prefactor of $\omega_R$ is $200$~MHz in typical dots, see Eq.~\eqref{eq:Rabi_new-SQD}.
}
\end{figure}

When the magnetic field is aligned to the $x,y,z$ directions, these terms do not induce spin transitions, and only modulate the qubit energy. However, at arbitrary orientations of $\textbf{B}$, Rabi oscillations can be induced by making use of the tunable anisotropy of the $g$-factor~\cite{doi:10.1126/science.1080880,PhysRevLett.120.137702,doi:10.1063/1.4858959,PhysRevApplied.16.054034,VenitucciElectricalmanipulationsemiconductor2018}.
In particular, when the  qubit energy $\hbar\omega_B=\mu_B\sqrt{g_{xx}^2B_x^2+g_{yy}^2B_y^2+g_{zz}^2B_z^2}$ is at resonance with $\omega_D$, the Rabi frequency induced by $E_x^{ac}$ is 
\begin{equation}
\label{eq_rabi-gmod}
\frac{\omega_R}{2\pi}=\frac{\omega_D}{2\pi} \frac{eE_x^{ac}R}{2\hbar\omega_\varphi}\left|\frac{\delta \underline{g}^x\cdot \textbf{B}}{\hbar\omega_B/\mu_B}-\frac{(\delta \underline{g}^x\cdot \textbf{B})\cdot(\underline{g}\cdot \textbf{B})}{\hbar^3\omega_B^3/\mu_B^3} (\underline{g}\cdot \textbf{B}) \right| \ .
\end{equation}

In Fig.~\ref{fig:drive_SQD_gmod}, we analyze the dependence of this driving mechanism on the direction of the magnetic field for different values of the dc  field $E_x$.
We consider here an arbitrary field $\textbf{B}=B (\cos(\theta_B)\sin(\varphi_B),\cos(\theta_B)\cos(\varphi_B),\sin(\theta_B))$, see Fig.~\ref{fig:sketch}.
We observe that $\omega_R=0$ when $\textbf{B}$ is aligned to a confinement axis, but in certain parameter regimes it becomes comparable to the values obtained by driving $E_y^{ac}$.
Comparing Fig.~\ref{fig:drive_SQD_gmod}a) to Fig.~\ref{fig:gfact_SQD}, we note that  when $\textbf{B}$ is slightly misaligned from the $z$-direction, a large $\omega_R$ can be reached  close to the value of $E_x$ where $g_{zz}$ vanishes; in particular, at $E_x=0.4\hbar\omega_\varphi/eR$, $g_{zz}\approx 1$.
Away from these sweet spots in both electric field and magnetic field direction, $\omega_R$ is strongly reduced, and the optimal direction of $\textbf{B}$ changes, see Fig.~\ref{fig:drive_SQD_gmod}b).
In  contrast, when the qubit is driven by $E_y^{ac}$, the optimal values of $\omega_R$ occur at the fixed direction $\textbf{B}\parallel y$ and persist in a wide range of $E_x$.
For this reason, the $E_y^{ac}$ driving is more convenient in experiments comprising chains of quantum dots in the CQW subjected to a global fixed $\textbf{B}$ field, and we focus on it in the following.

\subsection{Planar Ge curved quantum wells}
\label{sec:planar}
\begin{figure}
\centering
\includegraphics[width=0.45\textwidth]{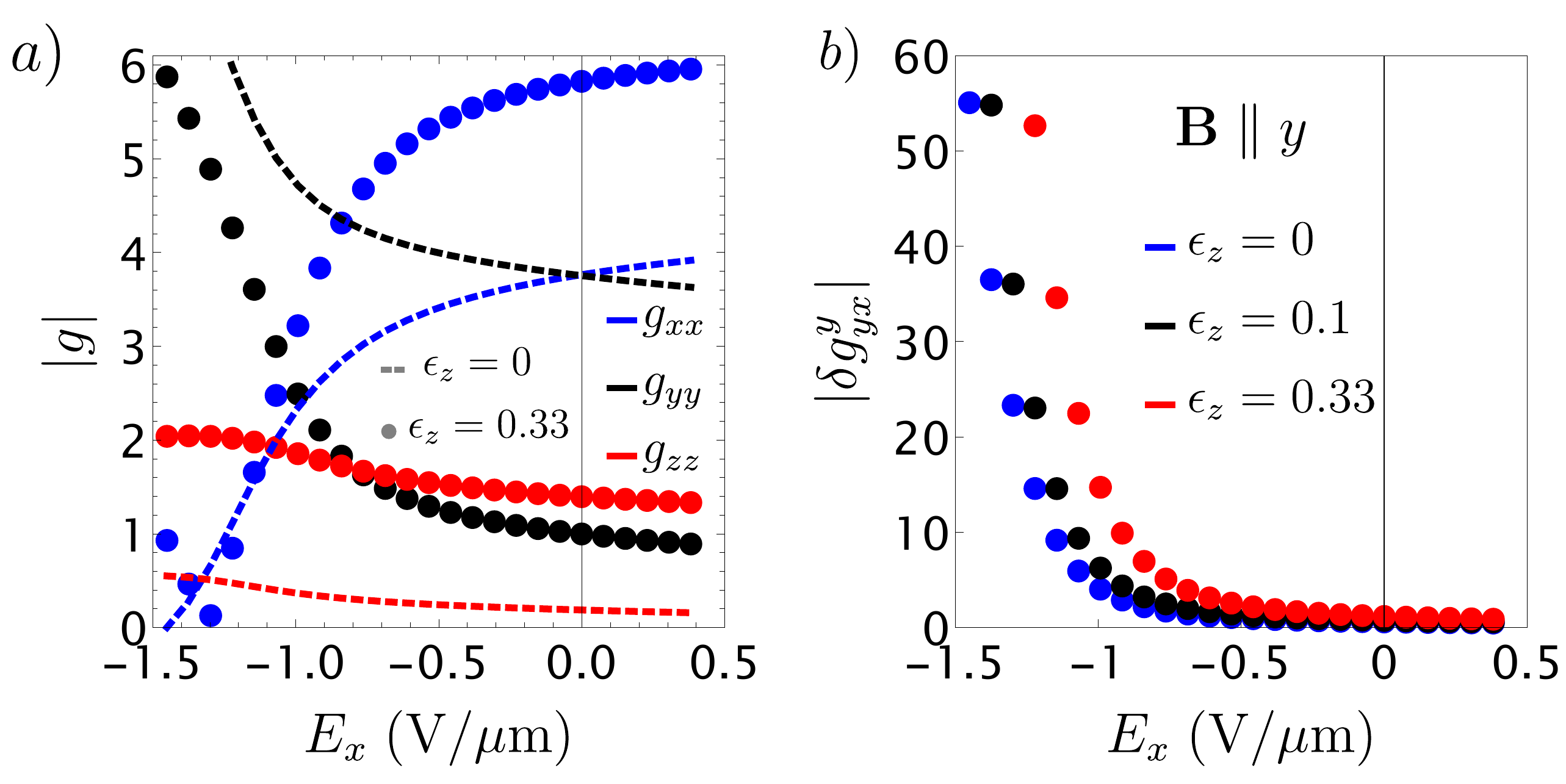}
\caption{\label{fig:drive_SQD_half}
Hole spin qubit in a planar thin shell quantum dot. In a), we show the $|g|$ factor as a function of the electric field applied perpendicular to the substrate and in b), we show the driving term $\delta g_{yx}^y$, obtained by an ac field $E_y(t)$ aligned to the magnetic field $\textbf{B}$. For typical experimental parameters the Rabi frequency is $\omega_R/2\pi\approx 200$~MHz$\times \delta g_{yx}^y/g_{yy}$.
For the plots, we consider $l_z=R=\tau/2=20$~nm, see Fig.~\ref{fig:sketch}b), and $\epsilon_r=\epsilon_c$. The values of $\epsilon_z$ are given in units of $\epsilon_c$.
}
\end{figure}

Similar qubits can be designed also in planar systems, providing a technologically competitive and scalable architecture for hole-based quantum computers. A possible example of a planar CQW  is sketched in Fig.~\ref{fig:sketch}b) and can be manufactured by growing  a Ge quantum well over a Si substrate.
Because the spin qubit proposed in Sec.~\ref{sec:SQD} works well at $E_x\approx\hbar\omega_\varphi/eR$, where the hole wavefunction is confined to the top half of the shell, see Fig.~\ref{fig:band}b), we expect an analogous behaviour in a spin qubit defined in planar CQWs. 

In Fig.~\ref{fig:drive_SQD_half}, we show the results of a numerical simulation of the Hamiltonian in Eq.~\eqref{eq:Hamiltonian} defined in the planar CQW sketched in Fig.~\ref{fig:sketch}b). In Fig.~\ref{fig:drive_SQD_half}a), we show the $g$-factor of a quantum dot of length $l_z=R$  as a function of the electric field $E_x$ perpendicular to the substrate.
We observe that when the magnetic field is perpendicular to the well, there is a critical electric field above which the behaviour of $g_{xx}$ and  $g_{yy}$ as function of $E_x$  resembles the  one shown in Fig.~\ref{fig:gfact_SQD}b). The position of this critical electric field depends on the longitudinal strain $\epsilon_z$ and is shifted to more negative values as $\epsilon_z$ increases.
Moreover, in analogy to  annular CQWs,  $g_{zz}$ is a rather flat function of the $E_x$ and it increases with $\epsilon_z$, approaching the high field limiting value $g_{zz}\approx 6 \kappa \tilde{\epsilon}_z\cos(\theta_z)$. These results indicate that qubits in planar CQWs can be as insensitive to charge noise as  qubits in annular CQWs.

In analogy to Fig.~\ref{fig:drive_SQD}b), we also find that these qubits can be driven fast by an ac electric field $E_y(t)$ applied parallel to the substrate. In particular,  we observe that the driving strength can easily exceed the GHz range and becomes even stronger than in annular CQWs as the dc electric field $E_x$ decreases. At lower values of $E_x$, however, the subband gap $E_g$ separating computational and non-computational states also decreases. In the range of parameters examined here $E_g\gtrsim 0.5$~meV. 

We note that at $E_x\lesssim -2.5$~V/$\mu$m, $E_g$ drops to zero and one obtains two degenerate quantum dots at the two edges of the quantum well. We envision protocols where these two quantum dots can be exchange-coupled and used to perform fast read-out schemes in analogy to the corner dots in Si FETs~\cite{Voisin2014,doi:10.1063/1.4950976,PhysRevX.10.041010}, but we do not investigate these intriguing possibilities here.

\section{Spin qubits in Long quantum dots}
\label{sec:Elongated-QDs}

\begin{figure}
\centering
\includegraphics[width=0.5\textwidth]{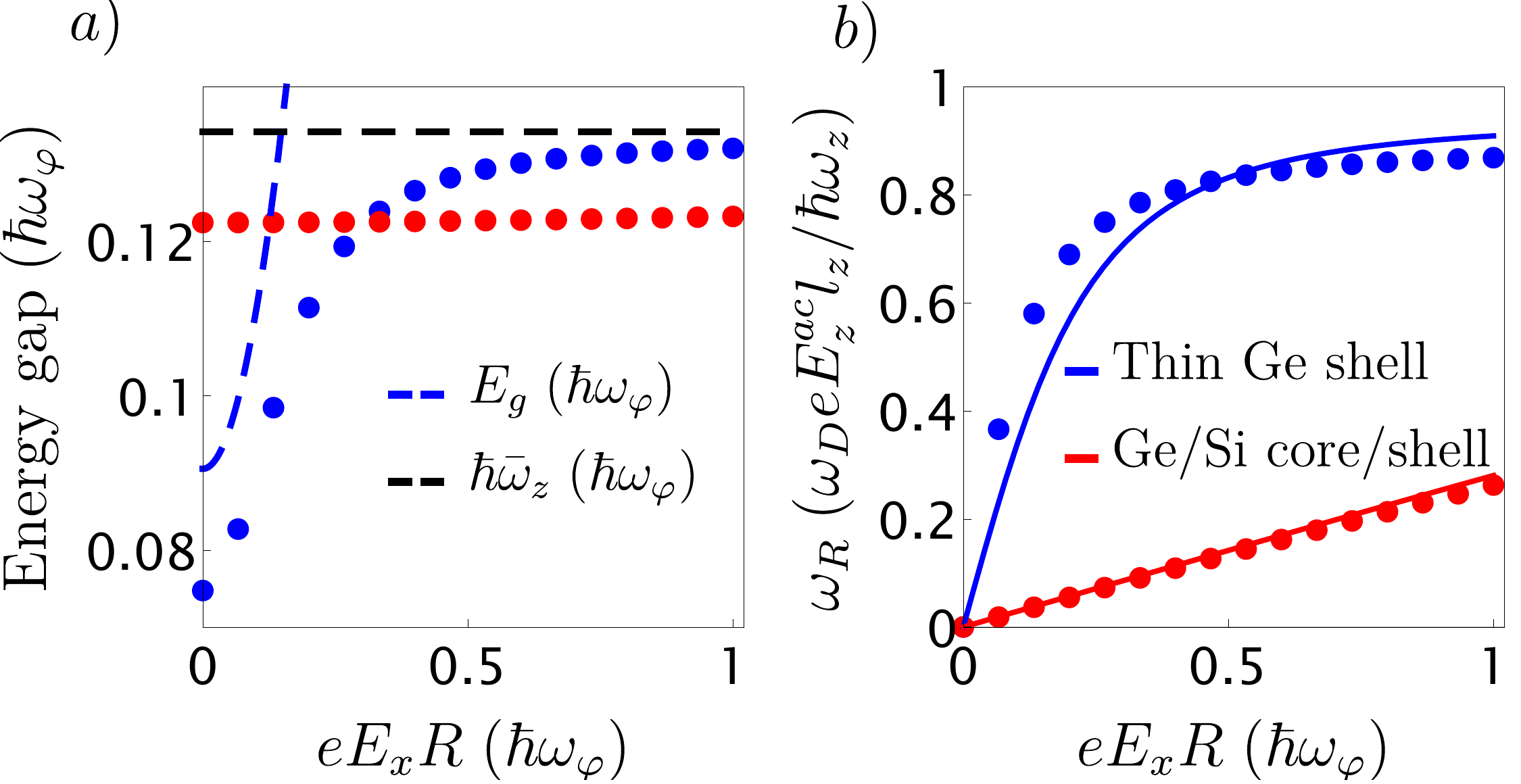}
\caption{\label{fig:SOI_EQD}  Electric field dependence of subband energy gap and Rabi frequency in long quantum dots. We compare thin Ge CQWs (blue) and Ge/Si core/shell nanowires (red). The lines are obtained by the effective theory in Eq.~\eqref{eq:long-dot-model}. The dots show a numerical simulation of a three-dimensional quantum dot obtained by discretizing  the Hamiltonian~\eqref{eq:Hamiltonian} with an annular (cylindrical) cross section, a harmonic potential along $z$, and with $\epsilon_r=3\epsilon_z=3\epsilon_c$ ($\epsilon_z=\pi^2\hbar\omega_\varphi/4$) for the CQW (core/shell nanowire). 
We use $\tau=R/2$ and $l_z=3R$. 
In a), we show the energy gap between the ground and first excited doublets. The dashed blue line is obtained by Eq.~\eqref{eq:gap-corrected} and the dashed black line shows $\bar{\omega}_z$. 
 In b) we show the Rabi frequency $\omega_R$ at resonance, see Eqs.~\eqref{eq:Rabi_EQD},~\eqref{eq:SOI-long}. For realistic experimental parameters $\omega_D/2\pi=5$~GHz, $E_z^{ac}=10$~mV/$\mu$m, $l_z=30$~nm and  $\hbar\omega_z=1.1$~meV, the prefactor of the Rabi frequency is $\omega_D eE_z^{ac}l_z/\hbar\omega_z\approx 1.4$~GHz. For the Ge/Si core/shell nanowire, we use the mass ${m}_z=0.06 m$ and the SOI velocity $v=2\times 0.15\times 6.5 eE_xR/\Delta_\text{c/s}$, with gap $\Delta_\text{c/s}= 0.5 \epsilon_z$.   }
\end{figure}

When  $l_z\gg R$, the longitudinal confinement energy $\hbar\omega_z=\hbar^2\gamma_1/ml_z^2$ is much smaller than angular momentum quantization energy $\hbar\omega_\varphi$. However, because of longitudinal strain, $\hbar\omega_z$ can still be comparable to the SOI energy $\Delta$, that is by a factor $3\tilde{\kappa}/2\gamma_1\sim 0.5-0.15$ smaller than $\hbar\omega_\varphi$, see Eq.~\eqref{eq:delta}. 
For this reason, a simple description of the system for moderate electric fields is obtained by projecting the effective theory in Eqs.~\eqref{eq:effective_H} and~\eqref{eq:E-field_H} onto the low-energy states $|g_{1,2}^{l=1}\rangle$ and $|e_{1,2}^{l=1}\rangle$ in Eq.~\eqref{eq:eigenstates_pz0}, resulting in the one-dimensional Hamiltonian
\begin{equation}
\label{eq:long-dot-model}
H_\text{LD}=\frac{p_z^2}{2m_z}-\frac{\Delta}{2} \lambda_z-\frac{eE_xR}{2} \lambda_x+v_z p_z \lambda_x\sigma_y+\frac{m_z\bar{\omega}_z^2}{2}z^2 \ ,
\end{equation}
where $\lambda_{x,y,z}$ and $\sigma_{x,y,z}$ are Pauli matrices acting on the orbital and pseudospin subspaces, respectively.

In this regime, the Ge CQW mimics a Ge/Si core/shell nanowire, and Eq.~\eqref{eq:long-dot-model} is qualitatively analogous to the well-known model discussed in detail in Refs.~\cite{DRkloeffel1,DRkloeffel2,DRkloeffel3,adelsberger2021hole}.
The direct Rashba SOI velocity~\cite{DRkloeffel3} 
\begin{equation}
\label{eq:SOI-long}
v= \frac{eE_xR}{\sqrt{\Delta^2+(eE_xR)^2}}v_z  \ ,
\end{equation}
is particularly important for spin qubits. In fact,  in  EDSR experiments~\cite{PhysRevB.74.165319,froning2020ultrafast,bosco2021squeezed} where an ac field $E_z^{ac}\cos(\omega_D t)$ is applied along the quantum well, $v$ is directly related to the Rabi frequency $\omega_R$  by
\begin{equation}
\label{eq:Rabi_EQD}
\frac{\omega_R}{2\pi}= \frac{\omega_D}{2\pi}  \frac{v eE_z^{ac}}{\hbar\bar{\omega}_z^2} \ .
\end{equation}
Moreover, the subband energy gap $\Delta$ is renormalized by $E_x$ and by the SOI velocity $v_z$ and it is approximately corrected as
\begin{equation}
\label{eq:gap-corrected}
\Delta\to  E_g\approx e^{-\frac{m_z^2 v_z^2 \bar{l}_z^2}{\hbar^2[\Delta^2+(eE_xR)^2]}}  \sqrt{\Delta^2+(eE_xR)^2} \ .
\end{equation} 
Eq.~\eqref{eq:SOI-long} and the square root in Eq.~\eqref{eq:gap-corrected} are obtained by rewriting Eq.~\eqref{eq:long-dot-model} in the basis that diagonalizes $H_\text{LD}$ at $p_z=0$.  The overall Gaussian suppression of the subband gap is most relevant at weak electric field values. It can be derived by first removing the direct couplings between ground state and first excited states by performing a spin and orbital dependent shift  $e^{- i m_z v_z z \lambda_x\sigma_y/ \hbar \sqrt{\Delta^2+(eE_xR)^2}}$ and then by averaging the resulting potential $ \sqrt{\Delta^2+(eE_xR)^2} \cos\left[2 m_z v_z z/\hbar \sqrt{\Delta^2+(eE_xR)^2}\right]$ over the harmonic oscillator ground state. This correction closely resembles the SOI-induced $g$-factor renormalization in nanowires with a large SOI~\cite{PhysRevB.98.165403, adelsberger2021hole,DRkloeffel2, PhysRevLett.127.190501}, and in the short dot case  it approaches Eq.~\eqref{eq:Eg_SQD}.

As shown in Fig.~\ref{fig:SOI_EQD}, the simple approximate Eqs.~\eqref{eq:SOI-long} and~\eqref{eq:gap-corrected} give a reasonable agreement to a complete three-dimensional simulation of a quantum dot even when the dots is moderately short and $E_g\sim \hbar\bar{\omega}_z$.

There are some noteworthy quantitative differences between CQWs and core/shell nanowires that  significantly impact the performance of the spin qubit; these differences are highlighted in Fig.~\ref{fig:SOI_EQD}. 
First, as discussed in Sec.~\ref{sec:Effective-theory}, because of strain the energy gap $\Delta$ is much smaller in a CQW than in core/shell nanowires where $\Delta_\text{c/s}\approx 0.5 \epsilon_z\sim 10$~meV. For this reason,  in moderately long quantum dots defined in a CQW the energy gap to the first excited doublets is $E_g$ at $E_x=0$ and it approaches $\hbar\omega_z$ only at finite values of  $E_x$.
Moreover, the dipole energy $-e E_x R\lambda_x/2$ in Eq.~\eqref{eq:long-dot-model} is 3.3 times larger in a CQW than in a core/shell nanowire, where this term is $-0.15 e E_x R\lambda_x$~\cite{DRkloeffel2}.
The significantly different ratio of  dipole energy and  subband gap strongly impacts the dependence of the Rabi frequency $\omega_R$ on $E_x$, see Fig.~\ref{fig:SOI_EQD}b). In fact, in a CQW the optimal direct Rashba SOI is obtained at values of $E_x$ that are $\sim 10$ times smaller than in core/shell nanowires, and $v$ remains constant over a wide range of $E_x$, enabling ultrafast qubit operations at low power even at small $E_x$.

We also remark that, as discussed in Sec.~\ref{sec:decoh}, in strained CQW at sufficiently large $E_x$, when the magnetic field is applied along the quantum well, the $g$-factor is to good approximation independent of $E_x$, suppressing the sensitivity of the spin to charge noise and  significantly boosting the coherence time and  fidelity of these qubits.

\section{Strong Spin-photon coupling}
\label{sec:spin-photon}
The large dipole moment of these quantum dots and their large SOI  makes this architecture optimal to strongly couple the hole spin qubit to a superconducting resonators. 
One viable approach is to couple a long quantum dot to a high-impedance resonator  by shaking the dot along the smooth confinement direction. In analogy to Ge/Si core/shell nanowires~\cite{DRkloeffel2,bosco2022fully}, this approach enables a strong spin-photon interaction in a single quantum dot, and it is especially appealing in our system when the magnetic field is applied along the $z$-direction, where the qubit is insensitive to charge noise, see Sec.~\ref{sec:decoh}.
In this case,  the large direct Rashba SOI linear in momentum $p_z$, see Eqs.~\eqref{eq:long-dot-model} and~\eqref{eq:SOI-long}, enhances the strength of the spin-photon interaction compared to alternative proposals based on cubic SOI~\cite{PhysRevB.102.205412}, potentially resulting in orders of magnitude larger coupling strengths~\cite{bosco2021squeezed}.
 However, this approach  requires a plunger gate misaligned from the center of the quantum dot, reducing the geometric lever arm between the electrode and the dot, and potentially risking the screening of the driving gate  by the electrode defining the dot.

In contrast,  we now focus here on a different setup that makes use of the alternative driving mechanism discussed in Sec.~\ref{sec:driving}, where a short quantum dot is shaken in the angular direction. Because this approach relies on a driving field applied in $y$-direction, perpendicular to the smooth confinement, the plunger electrode can be aligned to the center of the dot, enhancing the lever arm and potentially enabling higher coupling strengths.  While we now restrict our analysis to short quantum dots in annular CQWs, we emphasize that the results shown here are  valid  also for planar CQWs, see Sec.~\ref{sec:planar}.
We remark that our system only requires a \textit{single short quantum dot}, in contrast to different approaches in hole systems where the dipole moment is enlarged by delocalizing the hole over more  dots~\cite{PhysRevResearch.3.013194}.

State-of-the-art high impedance resonators can be made rather resilient against small magnetic fields, with quality factors $Q\sim 10^5$ at the small magnetic fields  $B\lesssim 0.5$~T considered here, and at the same time they reach rather high values of zero-point-fluctuation potential $V_\text{ZPF}=e\omega_C\sqrt{\hbar Z}\sim 10-100$~$\mu$eV~\cite{PhysRevApplied.5.044004,PhysRevApplied.11.044014,Grunhaupt2019,Maleeva2018,PhysRevLett.121.117001}.  Here,  $Z\sim 1-10$~k$\Omega$ is the characteristic impedance of a cavity with resonant frequency $\omega_C/2\pi\approx 5$~GHz.

The Hamiltonian describing this cavity is $H_C=\hbar\omega_C a^\dagger a$, where $a$ and $a^\dagger$ are bosonic ladder operators annihilating and creating a microwave photon in the resonator. At the antinode the quantized electric potential of a single boson is $\hat{V}=V_\text{ZPF} (a^\dagger+a)$~\cite{girvin2014circuit,RevModPhys.93.025005}.
If the plunger electrode is connected to the antinode of the resonator instead of an external power source, then the qubit Hamiltonian in Eq.~\eqref{eq:QD-effective-theory} is still valid, while the ac drive is replaced as $eE_y(t)R\to \alpha V_\text{ZPF} (a^\dagger +a)$, where $\alpha$ is the lever arm of the electrode. We neglect here the variations of the quantum dot size caused by the gate and assume that the electrode  only produces an electric field $E_y$. 

To maximize the qubit-resonator interactions, we consider $\textbf{B}\parallel y$, resulting in the coupling Hamiltonian $H_\text{int}=\nu (a^\dagger+a)\sigma_x$, with interaction strength
\begin{equation}
\frac{\nu}{2\pi}= \frac{\omega_B}{2\pi}  \frac{\alpha V_\text{ZPF}}{2\hbar\omega_\varphi}\frac{\delta g_{xy}^y}{g_{yy}} \ .
\end{equation}
By considering as in Eq.~\eqref{eq:Rabi_new-SQD} a strained  CQW with radius $R=20$~nm,  $\omega_B/2\pi= 5$~GHz and $E_x=0.13$~V/$\mu$m, such that $\delta g_{xy}^y/g_{yy}\approx 6$, $g_{yy}=1.35$ and $B=0.25$~mT, we find $\nu\approx 50$~MHz for the realistic values of lever arm $\alpha=0.4$ and $V_\text{ZPF}=  20$~$\mu$eV~\cite{PhysRevApplied.5.044004}.
This interaction strength is  comparable to that reported in charge qubits~\cite{PhysRevX.7.011030} and in spin qubits defined in multiple quantum dots~\cite{mi2018coherent,Landig2018}.
Moreover, we note that this system is well within the strong coupling regime. In fact, $\nu$ is about 40 times larger than the dephasing rate $1/T_2^*\approx 1.3$~MHz of the qubit, see Sec.~\ref{sec:decoh} and Fig.~\ref{fig:deph}, and three orders of magnitude larger than the decay rate of the photon in state-of-the-art cavities, $\omega_C/2\pi Q\approx 50 $~kHz.

We emphasize that the values of $\nu$ reported here can be further optimized in different ways.
Larger values of the ratio $\delta g_{xy}^y/g_{yy}\omega_\varphi$ can  be reached by tuning $E_x$, by increasing the radius $R$, and, in our device, also by optimizing the  electrostatic gate design to maximize the lever arm.
Moreover, a stronger coupling strength can be reached at larger cavity and qubit frequencies and higher impedances because $\nu\propto \omega_B\omega_C\sqrt{Z}$. By considering  $B=1$~T~\cite{PhysRevApplied.5.044004} and by reducing the length of resonator by $4$ times, one obtains more than an order of magnitude larger $\nu/2\pi\approx 800$~MHz at $\omega_B/2\pi\approx 20$~GHz, a frequency still compatible with  microwave technology~\cite{mills2021two,zwerver2021qubits}.
Moreover, resonators with a higher characteristic impedance, approaching the resistance quantum $25$~k$\Omega$, could also be conceived e.g. by using carbon nanotubes~\cite{doi:10.1063/1.4868868,1406008,Chudow2016} or quantum Hall edge states~\cite{PhysRevB.100.035416,PhysRevApplied.12.014030, PhysRevB.96.115407,PhysRevResearch.2.043383}, further enhancing the coupling strength. The latter approach is particularly appealing for our system, because in Ge/SiGe heterostructures well-developed quantum Hall plateaus have been recently observed at magnetic fields below 1~T~\cite{lodari2021lightly}.

For these reasons,  in our devices strong spin-photon couplings with interaction strength exceeding a few hundreds of MHz are realistically achievable with current state-of-the-art technology, opening up new possibilities for entangling distant qubits, as well as for high fidelity single-shot readout schemes.

\section{Conclusions}

In conclusion, in this work we discussed annular and planar curved quantum wells, focusing on their application for spin-based quantum information processing. 
This architecture takes full advantage of the large SOI of hole nanostructures and the curvature of the cross-section enhances the electric dipole moment of the system, and guarantees that the maximal value of the SOI is reached at low values of the externally applied electric field. 

We presented a detailed model of these devices, discussing several possible implementations and highlighting their key features and their differences from current state-of-the-art hole spin qubits, including their peculiar response to strain and electric field. 
Strikingly, in a wide range of electric fields, CQWs are to good approximation  insensitive to charge noise, a critical issue in current devices, enabling ultrafast high coherent qubit gates at low power, and pushing hole spin qubits towards new speed and coherence standards.

We also find that in CQWs ultrafast operations can be realized in short quantum dots, with ac driving fields perpendicular to the well. This feature  enables a strong interaction between a hole spin confined in a single quantum dot and  microwave photons, with interaction strengths that  can realistically exceed a few hundreds of MHz with current technology. CQWs can thus relax the many technological constraints and challenges to reach the strong hole spin-photon coupling regime, and will constitute an effective building block  to scale up the next generation of quantum processors.

\begin{acknowledgments}
We thank C. Adelsberger, B. Hetenyi, and H. Legg for useful discussions, and T. Patlatiuk and G. Katsaros for valuable comments and feedback on the manuscript.
We are also grateful to G. Gadea Diez and I. Zardo for drawing our attention towards curved quantum wells.
This work was supported as a part of NCCR SPIN funded by the Swiss National Science Foundation (grant number 51NF40-180604).
\end{acknowledgments}

\appendix

\section{Strain Hamiltonian}
\label{app:strain}

We present here a more detailed derivation of the strain Hamiltonian in Eq.~\eqref{eq:strain}.
In general, the effect of strain on the valence band of Si and Ge is  modelled by the isotropic Bir-Pikus (BP) Hamiltonian~\cite{bir1974symmetry}
\begin{equation}
\label{eq:BP-Ham}
H_\text{BP}= b\sum_{i} \varepsilon_{ii}J_i^2+2 b \varepsilon_{xy}\{J_x,J_y\}+\text{c.p.} \ ,
\end{equation}
where c.p. stands for cyclic permutations,  $b=-2.2$~eV, and $\varepsilon_{ij}$ are the elements of the strain tensor.

The strain tensor in the Ge quantum well can be accurately estimated by using classical linear elasticity theory~\cite{landaulifshitz_elasticity}, in analogy to~\cite{PhysRevB.90.115419,doi:10.1063/1.3207838,PhysRevB.50.10970,doi:10.1063/1.2337110,doi:10.1063/1.1601686}.
We consider an infinitely long annular CQW with the cross-section sketched in Fig.~\ref{fig:sketch}a). Following Ref.~\cite{PhysRevB.90.115419}, we assume that Si and Ge are isotropic elastic media, where the linear relation between the stress tensor $\sigma_{ij}$ and the strain tensor $\varepsilon_{ij}$ is parametrized by two Lamé parameters $\lambda$ and $\mu$. The static displacement field $\textbf{u}$  is related to the strain tensor by $\varepsilon_{ij}=(\partial_i u_j+\partial_j u_i)/2$.
In the absence of body forces, the displacement field  $\textbf{u}$ in the material $n=\{\text{Si, Ge}\}$ is static when~\cite{landaulifshitz_elasticity}
\begin{equation}
\mu_n \nabla^2 \textbf{u}^n+ \left(\mu_n+\lambda_n\right)\nabla\left(\nabla\cdot \textbf{u}^n\right)=0 \ .
\end{equation}

The boundary conditions of this system of equations are extensively discussed e.g. in Refs.~\cite{PhysRevB.90.115419,doi:10.1063/1.3207838,doi:10.1063/1.2337110,PhysRevB.50.10970} and they include the absence of forces at the interfaces and the condition of pseudomorphic growth $\textbf{t}\cdot \textbf{l}_n=\textbf{t}\cdot\textbf{l}_m$, where $\textbf{t}$ is a tangent vector to the interface and the distorted lattice vectors of the material $n,m$ are
\begin{equation}
\textbf{l}_{n,m}= a_{n,m} \sum_i  \textbf{e}_i\sum_j\left( \delta_{ij}+\partial_i u_j^{n,m}\right) \; ,
\end{equation}
where $a_\text{Si}=0.543$~nm and $a_\text{Ge}=0.566$~nm are the lattice constants of Si and Ge, respectively.
Finally, a stable strain configuration  minimizes the total elastic energy $U=\frac{1}{2}\sum_n\int d\textbf{r}\left(\lambda_n \text{Tr}(\varepsilon^n)^2+2\mu_n\sum_{ij}[\varepsilon^n_{ij}]^2 \right) $.

By working in cylindrical coordinates, we find that in the Ge quantum well the strain tensor is diagonal and its diagonal elements read 
\begin{equation}
\varepsilon_{rr,\varphi\varphi}= \frac{1}{2}\left(c_1\pm \frac{R_1^2}{r^2}c_2\right) \ , \ \varepsilon_{zz}=\frac{c_3}{2} \ ,
\end{equation}
with $c_i$ being dimensionless constants dependent on the design of the quantum well.
Combining this result with Eq.~\eqref{eq:BP-Ham}, the BP Hamiltonian reduces to
\begin{equation}
\label{eq:strain-1}
H_\text{BP}=J_z^2\epsilon_z- J_r^2\epsilon_r+\left(J_r^2+\frac{1}{2}J_z^2-\frac{15}{8}\right)\left(1-\frac{R^2}{r^2}\right) \epsilon_r  \ .
\end{equation}

Neglecting the small corrections arising form the difference in the Lamé parameters of Si and Ge, the energies $\epsilon_{r}$ and $\epsilon_z$ can be compactly written as
\begin{subequations}
\label{eq_strain_pars-1}
\begin{align}
\epsilon_r & =  \frac{R_1^2}{R^2} |b| c_2 \approx \left(1-\frac{\tau}{2R}\right)^2|b| \varepsilon_0 \ , \\
\epsilon_z& =\frac{|b|}{2}\left(c_1-\frac{R_1^2}{R^2}c_2-c_3\right)  \approx
\left(\frac{1}{2}-\frac{\tau}{8R}-\frac{R^2}{R_3^2}\right) \frac{\tau}{R}|b|\varepsilon_0 \ , \\
\varepsilon_{0}&=\left(1+\frac{2 \lambda}{\lambda+2 \mu}\right)\varepsilon_{\parallel}\approx 1.6 \varepsilon_\parallel \ .
\end{align}
\end{subequations} 
The expressions obtained by including the differences of Lamè parameters of Si and Ge are lengthy and we do not report them here. However, we show them with solid lines in Fig.~\ref{fig:strain}, and we find that Eqs.~\eqref{eq_strain_pars-1} are rather accurate and nicely reproduce the  more general solutions.

Compared to Eq.~\eqref{eq:strain}, Eq.~\eqref{eq:strain-1} presents an additional, inhomogeneous radial strain component, that is proportional to $1-R^2/r^2$. This term is completely off-diagonal when the quantization axis is along the $z$-direction and it mixes HH and LH with opposite spins. 
We note that the matrix elements of the operator $1-R^2/r^2$ in the basis states in Eq.~\eqref{eq:basis-states} are $\propto\tau$ and they mix states that are separated by an energy gap $\sim \epsilon_c\propto\tau^{-2}$. Consequently, the corrections to the model caused by the inhomogeneous strain scale as $\propto \tau^3$, and because in this work we focus on thin quantum wells, we neglect them in the main text.

\section{Hamiltonian in the rotated basis}
\label{sec:LK-BP-cyl}
We report here the explicit expressions of the $6\times 6$ LK, magnetic and BP Hamiltonians including HHs, LHs and spin-orbit split-off holes (SOHs)~\cite{WinklerSpinOrbitCoupling2003}, after the transformation $U=e^{-i(J_3\oplus\sigma_3)\varphi}e^{-i(J_2\oplus\sigma_2)\pi/2}$, that generalizes the rotation discussed in the main text when the SOHs are included. 
To simplify the notation, for each Hamiltonian $H_i$, we introduce the decomposition 
\begin{equation}
U^\dagger H_i U= \left(
\begin{array}{cc}
H_i^{4} & H_i^{4,2}  \\
(H_i^{4,2})^\dagger & H_i^2   
\end{array}
\right) \ .
\end{equation}
We also write the equations assuming that the momentum operator always acts first on the wavefunction, e.g. $ r^{-1} k_r\equiv k_r/r \neq k_r r^{-1} $; we also use here the wavevector operators $k_{\varphi,z}=-i\partial_{\varphi,z}$ and $k_{r}=-i (\partial_r+1/2r)$ instead of the momenta $p_i=\hbar k_i$.

Introducing the quantities $\gamma_\pm=\gamma_1\pm\gamma_s$, and $ k_\pm=\frac{k_\varphi}{r}\pm i k_z $ and the SOHs gap $\Delta_\text{S}$~\cite{WinklerSpinOrbitCoupling2003}, we find  that in the  rotated basis the LK Hamiltonian reads
\begin{widetext}
\begin{subequations}
\label{eq:H-rotated}
\begin{align}
& H_\text{LK}^4 = \frac{\hbar^2}{m}\left(
\begin{array}{cccc}
 \frac{1}{2} k_r^2 \left(\gamma _1-2 \gamma _s\right)+\frac{2 \gamma _1-\gamma _s}{8 r^2} & 0 & \frac{\sqrt{3} \left(2 \gamma _1-\gamma _s\right)}{8 r^2}+\frac{i \sqrt{3} k_r \gamma _s}{2 r} & 0 \\
 0 &  \frac{1}{2} k_r^2 \left(\gamma _1+2 \gamma _s\right)+\frac{6 \gamma _1+9 \gamma _s}{8 r^2} & 0 & \frac{\sqrt{3} \left(2 \gamma _1+3 \gamma _s\right)}{8 r^2}-\frac{i \sqrt{3} k_r \gamma _s}{2 r} \\
 \frac{\sqrt{3} \left(2 \gamma _1+3 \gamma _s\right)}{8 r^2}-\frac{i \sqrt{3} k_r \gamma _s}{2 r} & 0 &  \frac{1}{2} k_r^2 \left(\gamma _1+2 \gamma _s\right)+\frac{6 \gamma _1+9 \gamma _s}{8 r^2} & 0 \\
 0 & \frac{\sqrt{3} \left(2 \gamma _1-\gamma _s\right)}{8 r^2}+\frac{i \sqrt{3} k_r \gamma _s}{2 r} & 0 & \frac{1}{2} k_r^2 \left(\gamma _1-2 \gamma _s\right)+\frac{2 \gamma _1-\gamma_s}{8 r^2} \\
\end{array}
\right)  \\
 &+\frac{\hbar^2}{m}\left(
\begin{array}{cccc}
 \frac{1}{2} \gamma _+ k_- k_+ & \frac{\sqrt{3} \left(\gamma _- k_++k_- \left(\gamma _++4 i r k_r \gamma _s\right)\right)}{4 r} & \frac{1}{2} \sqrt{3} k_-^2 \gamma _s & \frac{3 k_- \gamma _s}{2 r} \\
 \frac{\sqrt{3} \left(\gamma _+ k_-+k_+ \left(\gamma _1+\left(3-4 i r k_r\right) \gamma _s\right)\right)}{4 r} & \frac{1}{2} \gamma _- k_- k_+ & \frac{\gamma _1 \left(k_-+k_+\right)+\left(2 k_--k_+\right) \gamma _s}{2 r} & \frac{1}{2} \sqrt{3} k_-^2 \gamma _s \\
 \frac{1}{2} \sqrt{3} k_+^2 \gamma _s & \frac{\gamma _1 \left(k_-+k_+\right)-\left(k_--2 k_+\right) \gamma _s}{2 r} & \frac{1}{2} \gamma _- k_- k_+ & \frac{\sqrt{3} \left(\gamma _+ k_++k_- \left(\gamma _1+\left(3-4 i r k_r\right) \gamma _s\right)\right)}{4 r} \\
 \frac{3 k_+ \gamma _s}{2 r} & \frac{1}{2} \sqrt{3} k_+^2 \gamma _s & \frac{\sqrt{3} \left(\gamma _- k_-+k_+ \left(\gamma _++4 i r k_r \gamma _s\right)\right)}{4 r} & \frac{1}{2} \gamma _+ k_- k_+ \\
\end{array}
\right) \ , \\
& H_\text{LK}^{4,2} = \frac{\hbar^2\gamma_s}{2\sqrt{2}mr}\left(
\begin{array}{cc}
 \sqrt{3} \left(k_+-2 i k_- r k_r\right) & i \sqrt{3} \left(k_r+2 i k_-^2 r\right) \\
 2 k_- k_+ r-k_r \left(4 r k_r+3 i\right) & -2 k_++k_- \left(-5+6 i r k_r\right) \\
 2 k_-+k_+ \left(5-6 i r k_r\right) & -2 k_- k_+ r+k_r \left(4 r k_r+3 i\right) \\
 \sqrt{3} \left(2 k_+^2 r-i k_r\right) & -\sqrt{3} \left(k_--2 i k_+ r k_r\right) \\
\end{array}
\right) \ , \ H_\text{LK}^2=\frac{\hbar^2\gamma_1}{2m}\left[ k_+ k_-+k_r^2+\frac{k_\varphi}{r}\sigma_1\right]+ \Delta_\text{S} \ .
\end{align}
\end{subequations}
\end{widetext}
We note that to explicitly verify that $H_\text{LK}$ is a hermitian operator,  one needs  the   relation $[k_r,1/r]=i/r^2$.

We now report the magnetic Hamiltonian $H_\textbf{B}$ which includes both the Zeeman and  orbital magnetic field effects. By using the isotropic LK Hamiltonian and the gauge $\textbf{A}= (B_yz-B_z y/2,-B_xz+B_z x/2,0)$, one obtains to linear order in $\textbf{B}$
\begin{multline}
H_\textbf{B}=\mu_BB_z h_z+ \mu_B[B_x\cos(\varphi)+B_y\sin(\varphi)] h_+ \\ + \mu_B[B_x\sin(\varphi)-B_y\cos(\varphi)] h_- \ ,
\end{multline}
where the dimensionless matrices $h_i$ are given by
%
\begin{widetext}
\begin{align}
h_z^4&=\left(
\begin{array}{cccc}
 \gamma _+ k_{\varphi } & \frac{\sqrt{3}}{2}  \left[\gamma _1-2 (\kappa - i r k_r \gamma _s- \gamma _s)\right] & \sqrt{3} k_- r \gamma _s & \frac{3 \gamma _s}{2} \\
 \frac{\sqrt{3}}{2}  \left[\gamma _1-2( \kappa + i r k_r \gamma _s)\right] & \gamma _- k_{\varphi } & \gamma _1-2 \kappa +\frac{\gamma _s}{2} & \sqrt{3} k_- r \gamma _s \\
 \sqrt{3} k_+ r \gamma _s & \gamma _1-2 \kappa +\frac{\gamma _s}{2} & \gamma _- k_{\varphi } & \frac{\sqrt{3}}{2} \left[\gamma _1-2 (\kappa + i r k_r \gamma _s)\right] \\
 \frac{3 \gamma _s}{2} & \sqrt{3} k_+ r \gamma _s & \frac{\sqrt{3}}{2} \left[\gamma _1-2( \kappa - i r k_r \gamma _s- \gamma _s)\right] & \gamma _+ k_{\varphi } \\
\end{array}
\right) \ , \\
h_+^4&=\left(
\begin{array}{cccc}
 3 \kappa -\frac{2 \gamma _+ z k_{\varphi }}{r} & -\frac{\sqrt{3} z \left(\gamma _1+2 i r k_r \gamma _s+2 \gamma _s\right)}{r} & \frac{2 i \sqrt{3} \gamma _s \left(r \left\{ z, k_z \right\}+i z k_{\varphi }\right)}{r} & -\frac{3 z \gamma _s}{r} \\
 -\frac{\sqrt{3} z \left(\gamma _1-2 i r k_r \gamma _s\right)}{r} & \kappa -\frac{2 \gamma _- z k_{\varphi }}{r} & -\frac{z \left(2 \gamma _1+\gamma _s\right)}{r} & \frac{2 i \sqrt{3} \gamma _s \left(r \left\{ z, k_z \right\}+i z k_{\varphi }\right)}{r} \\
 -\frac{2 \sqrt{3} \gamma _s \left(z k_{\varphi }+i r \left\{ z, k_z \right\}\right)}{r} & -\frac{z \left(2 \gamma _1+\gamma _s\right)}{r} & -\kappa -\frac{2 \gamma _- z k_{\varphi }}{r} & -\frac{\sqrt{3} z \left(\gamma _1-2 i r k_r \gamma _s\right)}{r} \\
 -\frac{3 z \gamma _s}{r} & -\frac{2 \sqrt{3} \gamma _s \left(z k_{\varphi }+i r \left\{ z, k_z \right\}\right)}{r} & -\frac{\sqrt{3} z \left(\gamma _1+2 i r k_r \gamma _s+2 \gamma _s\right)}{r} & -3 \kappa -\frac{2 \gamma _+ z k_{\varphi }}{r} \\
\end{array}
\right) \ , \\
h_-^4&=\left(
\begin{array}{cccc}
 -\frac{z \left[\gamma _1 \left(2 r k_r+i\right)+\left(-4 r k_r+i\right) \gamma _s\right]}{r} & \frac{\sqrt{3} \left(-2 i z k_{\varphi } \gamma _s+i \kappa  r-2 r \left\{ z, k_z \right\} \gamma _s\right)}{r} & -\frac{2 i \sqrt{3} z \gamma _s}{r} & 0 \\
 \frac{i \sqrt{3} \left(2 z k_{\varphi } \gamma _s-\kappa  r+2 i r \left\{ z, k_z \right\} \gamma _s\right)}{r} & -\frac{z \left[\gamma _1 \left(2 r k_r+i\right)+\left(4 r k_r-i\right) \gamma _s\right]}{r} & 2 i \kappa  & 0 \\
 0 & -2 i \kappa  & -\frac{z \left[\gamma _1 \left(2 r k_r+i\right)+\left(4 r k_r-i\right) \gamma _s\right]}{r} & \frac{\sqrt{3} \left(2 i z k_{\varphi } \gamma _s+i \kappa  r+2 r \left\{ z, k_z \right\} \gamma _s\right)}{r} \\
 0 & -\frac{2 i \sqrt{3} z \gamma _s}{r} & \frac{\sqrt{3} \left(-2 i z k_{\varphi } \gamma _s-i \kappa  r+2 r \left\{ z, k_z \right\} \gamma _s\right)}{r} & -\frac{z \left[\gamma _1 \left(2 r k_r+i\right)+\left(-4 r k_r+i\right) \gamma _s\right]}{r} \\
\end{array}
\right) \ , \\
h_z^{4,2}&=\left(
\begin{array}{cc}
 \frac{1}{2} \sqrt{\frac{3}{2}} \left(2 \kappa -\gamma _s-2 i r k_r\gamma _s \right) & -\sqrt{6} k_- r \gamma _s \\
 \sqrt{2} r k_{\varphi } \gamma _s & \frac{2 \kappa -\gamma _s+6 i r k_r\gamma _s }{2 \sqrt{2}} \\
 \frac{-2 \kappa -6 i r k_r \gamma _s+\gamma _s}{2 \sqrt{2}} & -\sqrt{2} r k_{\varphi } \gamma _s \\
 \sqrt{6} k_+ r \gamma _s & \frac{1}{2} \sqrt{\frac{3}{2}} \left(-2 \kappa +2 i r k_r \gamma _s+\gamma _s\right) \\
\end{array}
\right) \ , \ h_z^2=\gamma _1 k_{\varphi } -\sigma_1(2\kappa-\gamma_1/2) \ ,\\
h_+^{4,2}&=\frac{1}{r} \left(
\begin{array}{cc}
 \sqrt{\frac{3}{2}} z \left(1+2 i r k_r\right) \gamma _s & 2 \sqrt{6} \gamma _s \left(z k_{\varphi }-i r \left\{ z, k_z \right\}\right) \\
 \sqrt{2} \left(\kappa  r-2 z k_{\varphi } \gamma _s\right) & \frac{z \left(1-6 i r k_r\right) \gamma _s}{\sqrt{2}} \\
 \frac{z \left(-1+6 i r k_r\right) \gamma _s}{\sqrt{2}} & \sqrt{2} \left(2 z k_{\varphi } \gamma _s+\kappa  r\right) \\
 -2 \sqrt{6} \gamma _s \left(z k_{\varphi }+i r \left\{ z, k_z \right\}\right) & \sqrt{\frac{3}{2}} z \left(-1-2 i r k_r\right) \gamma _s \\
\end{array}
\right) \ , \ h_+^2=-\left(
\begin{array}{cc}
 -2 \kappa +\frac{2 \gamma _1 z k_{\varphi }}{r} & \frac{\gamma _1 z}{r} \\
 \frac{\gamma _1 z}{r} & 2 \kappa +\frac{2 \gamma _1 z k_{\varphi }}{r} \\
\end{array}
\right) \ , \\
h_-^{4,2}&=\frac{\sqrt{3}}{\sqrt{2}r}\left(
\begin{array}{cc}
  2 i z k_{\varphi } \gamma _s-i \kappa  r+2 r \left\{ z, k_z \right\} \gamma _s & i  z \gamma _s \\
 \frac{z \left(8 r k_r+i\right) \gamma _s}{\sqrt{3}} & -\frac{6 i z k_{\varphi } \gamma _s+i \kappa  r+6 r \left\{ z, k_z \right\} \gamma _s}{\sqrt{3}} \\
 \frac{i \left(6 z k_{\varphi } \gamma _s-\kappa r+6 i r \left\{ z, k_z \right\} \gamma _s\right)}{\sqrt{3}} & -\frac{z \left(8 r k_r+i\right) \gamma _s}{\sqrt{3}} \\
 -i  z \gamma _s &  -2 i z k_{\varphi } \gamma _s-i \kappa  r+2 r \left\{ z, k_z \right\} \gamma _s \\
\end{array}
\right) \ , \ h_-^{2}=-\left(
\begin{array}{cc}
 \frac{\gamma _1 z \left(2 r k_r+i\right)}{r} & -2 i \kappa  \\
 2 i \kappa  & \frac{\gamma _1 z \left(2 r k_r+i\right)}{r} \\
\end{array}
\right) \ .
\end{align}
\end{widetext}
To explicitly verify the hermiticity of $H_\textbf{B}$, one needs  $[k_\varphi,\cos(\varphi)]=i\sin(\varphi)$ and $[k_\varphi,\sin(\varphi)]=-i\cos(\varphi)$.

Finally, neglecting the inhomogeneous component of strain, the $6\times 6$ rotated BP Hamiltonian  is
\begin{widetext}
\begin{equation}
H_\text{BP}=\left(
\begin{array}{cccccc}
 \frac{3}{4} \left(\epsilon _z-3 \epsilon _r\right) & 0 & \frac{\sqrt{3} \epsilon _z}{2} & 0 & 0 & -\sqrt{\frac{3}{2}} \epsilon _z \\
 0 & \frac{1}{4} \left(7 \epsilon _z-\epsilon _r\right) & 0 & \frac{\sqrt{3} \epsilon _z}{2} & -\frac{2 \epsilon _r+\epsilon _z}{\sqrt{2}} & 0 \\
 \frac{\sqrt{3} \epsilon _z}{2} & 0 & \frac{1}{4} \left(7 \epsilon _z-\epsilon _r\right) & 0 & 0 & \frac{2 \epsilon _r+\epsilon _z}{\sqrt{2}} \\
 0 & \frac{\sqrt{3} \epsilon _z}{2} & 0 & \frac{3}{4} \left(\epsilon _z-3 \epsilon _r\right) & \sqrt{\frac{3}{2}} \epsilon _z & 0 \\
 0 & -\frac{2 \epsilon _r+\epsilon _z}{\sqrt{2}} & 0 & \sqrt{\frac{3}{2}} \epsilon _z & \frac{5}{4} \left(\epsilon _z-\epsilon _r\right) & 0 \\
 -\sqrt{\frac{3}{2}} \epsilon _z & 0 & \frac{2 \epsilon _r+\epsilon _z}{\sqrt{2}} & 0 & 0 & \frac{5}{4}  \left(\epsilon _z-\epsilon _r\right) \\
\end{array}
\right) \ .
\end{equation}
\end{widetext}

\section{Split-off holes}
\label{app:SOHs}
\begin{figure}
\centering
\includegraphics[width=0.45\textwidth]{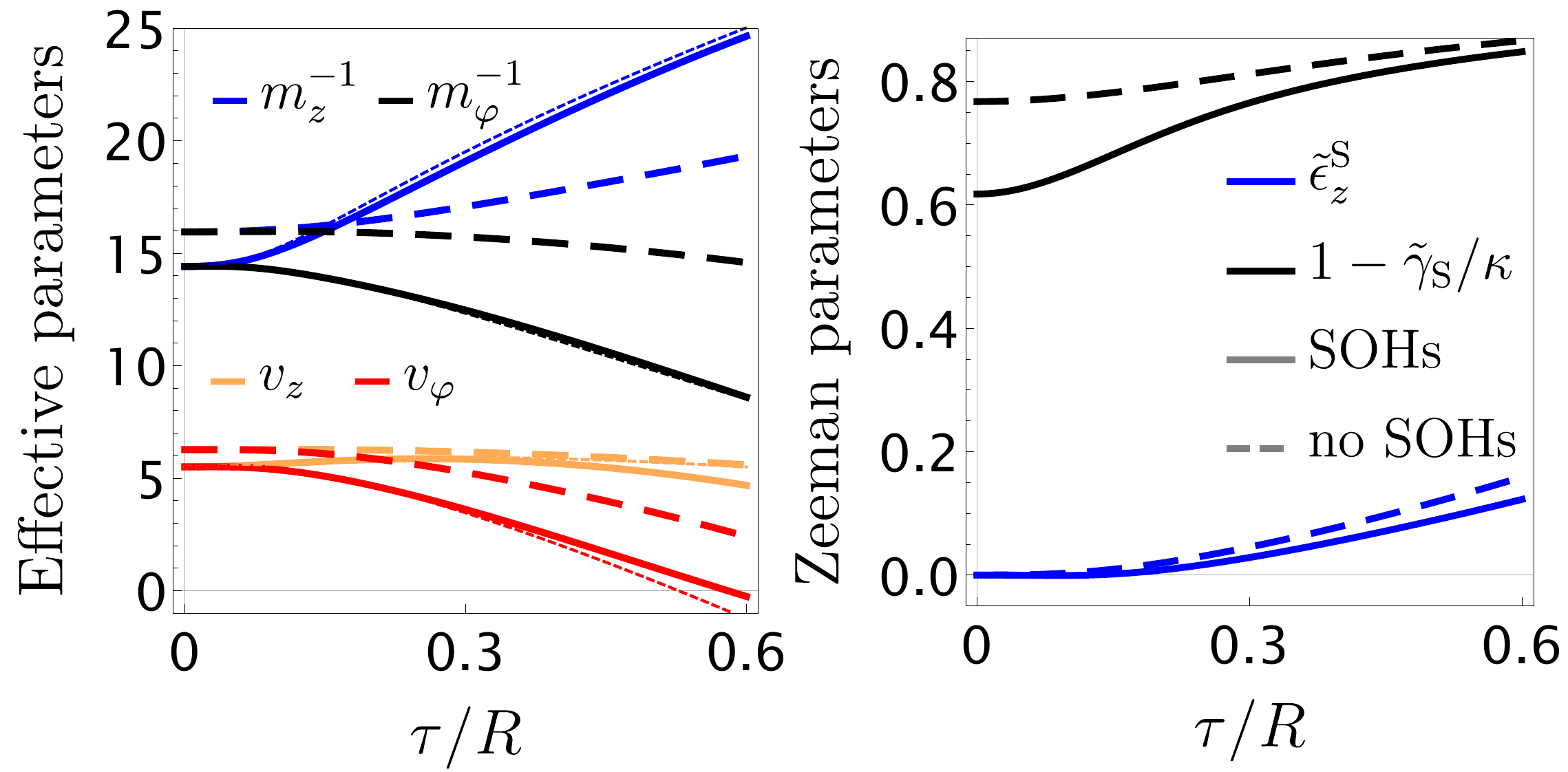}
\caption{\label{fig:pars_sohs} Effect of the SOHs on the parameters of the effective theory in Eq.~\eqref{eq:effective_H} of a thin Ge CQW. With dashed lines we show results obtained by neglecting the SOHs, see Eq.~\eqref{eq:pars}.  Solid lines and thin dotted lines represent results obtained by including corrections coming from $\Delta_\text{S}$. In particular thin dashed lines show the expression in Eq.~\eqref{eq:pars_SOH}, while solid lines show the complete results for the parameters to the $4^\text{th}$ order in Schrieffer-Wolff transformation. We consider here a thin quantum well with radius $R=20$~nm and vary the strain as a function of $\tau$ according to Eq.~\eqref{eq_strain_pars}. For simplicity, we consider here a thick outer shell with $R_3\gg R$.
The mass $m_{z,\varphi}=(1/m^*\pm 1/\delta m)^{-1}$ and the SOI $v_{z,\varphi}=v_-\pm v_+$  are shown in units of $m$ and $\hbar/mR$, respectively.
  }
\end{figure}

The valence band of cubic semiconductors comprises the HH and LH bands, degenerate at the $\Gamma$ point,  and a third band that is split-off by bulk SOI. In Ge these split-off holes (SOHs) are separated from the HHs and LHs by  $\Delta_\text{S}\approx 300$~meV~\cite{WinklerSpinOrbitCoupling2003}.
Our treatment can be generalized to include this additional band by the $6\times 6$ LK Hamiltonian reported in App.~\ref{sec:LK-BP-cyl}. The effective theory described by Eqs.~\eqref{eq:effective_H} and~\eqref{eq:Zeeman} is still qualitatively valid, but the parameters in Eq.~\eqref{eq:pars} are quantitatively modified  by $\Delta_\text{S}$. 
We find that the SOHs enhance the coupling of the hole spin to  photons in microwave resonators.

Because of $\Delta_\text{S}$, in the presence of a large longitudinal strain $\epsilon_z$ a second order perturbation theory does not suffice to accurately reproduce the spectrum. A good estimate of the effective parameters requires  terms up to 4$^\text{th}$ order in Schrieffer-Wolff perturbation theory.
The complete expressions of the parameters are lengthy and we do not report them here; however, by only keeping the most relevant terms, one simplify the expressions as 
\begin{subequations}
\label{eq:pars_SOH}
\begin{align}
\frac{m}{m^*}&\approx \gamma _1+\gamma _s -3 \tilde{\gamma}_\text{S} \ ,  \\ 
\frac{m}{\delta m}& \approx 3\gamma_s \left[\left(1+5\frac{\gamma_s\epsilon_c+\gamma_1\epsilon_r}{\gamma_1\epsilon_z}\tilde{\epsilon}_z^\Delta\right) \tilde{\epsilon}_z \right. \\
 & \  \ \ \ \ \ \left.+2\left(1+\frac{\gamma_s\epsilon_c+\gamma_1\epsilon_r}{\gamma_1\epsilon_z}\tilde{\epsilon}_z^\Delta\right)\tilde{\epsilon}_z^\Delta \right]  \ , \\
v_-&\approx \frac{3\hbar}{2m R}\left[\gamma_s-\tilde{\gamma}_\text{S}  -\left(\frac{\gamma_1}{2}+\gamma_s\right)\tilde{\epsilon}_z -\frac{3}{4}\left(\gamma_1+\gamma_s\right)\tilde{\epsilon}_z^\Delta\right]\ , \\
v_+ &\approx \frac{3\hbar}{4m R}\left[\gamma_1\tilde{\epsilon}_z +\frac{3}{2}\left(\gamma_1+\gamma_s\right)\tilde{\epsilon}_z^\Delta \right]\ ,
\end{align}
\end{subequations}
where we introduced the renormalized quantities
\begin{subequations}
\begin{align}
\tilde{\gamma}_\text{S}&= \tilde{\gamma}+\tilde{\gamma}_{\Delta}+ \frac{9\pi^2}{8\gamma_s}\tilde{\gamma}\tilde{\gamma}_{\Delta}\left[\frac{9\pi^2}{32\gamma_s}(\tilde{\gamma}+\tilde{\gamma}_{\Delta})-1\right] \ , \\
\tilde{\gamma}_{\Delta}&=\frac{128}{9 \pi^2}\frac{ \gamma _s }{2 + \gamma _1 \left(3 \epsilon _c+2\Delta_\text{S}+2 \epsilon _r+ \epsilon _z\right)/\gamma_s \epsilon _c } \ , \\
\tilde{\epsilon}_z^\Delta &= \frac{2 \epsilon _z}{\epsilon _z+2(\Delta_\text{S}+ \epsilon _r)+2 \gamma _s\epsilon _c/\gamma _1 } \ .
\end{align}
\end{subequations}

The Zeeman interactions in Eq.~\eqref{eq:Zeeman} are also quantitatively modified by $\Delta_\text{S}$ and they read
\begin{equation}
\label{eq:Zeeman_SOH}
\begin{split}
\frac{H_B}{3\kappa\mu_B}&=B_z \left[\left(
\tilde{\epsilon}_z^\text{S}+\frac{1}{2}\frac{\tilde{\kappa}_\text{S}}{\kappa}\right)\sigma_x+\frac{1}{3\kappa} \frac{m}{m_\varphi}p_\varphi   \right]\\
&+ \left[ B_x \cos(\varphi)+B_y\sin(\varphi)\right]\left[\left(1-\frac{\tilde{\gamma}_\text{S}}{\kappa}\right)\sigma_z-\frac{\tilde{\kappa}_\text{S}}{\kappa}\frac{z}{R} \sigma_x \right]\\
&+\left[ B_x \sin(\varphi)-B_y\cos(\varphi)\right]
\tilde{\epsilon}_z^\text{S}\sigma_y  \ ,
\end{split}
\end{equation}
where, in analogy to Eq.~\eqref{eq:delta}, we define $\tilde{\kappa}_{\text{S}}= {2mR }v_\varphi/{3\hbar}$. Here, $v_\varphi=v_--v_+$ and $m_\varphi=(1/m_*- 1/\delta m)^{-1}$ both include the corrections caused by $\Delta_\text{S}$.  The longitudinal strain modifies the Zeeman energy via the dimensionless parameter
\begin{multline}
\tilde{\epsilon}_z^\text{S}= \tilde{\epsilon}_z-\tilde{\epsilon}_z^\Delta+\frac{\gamma_s\epsilon_c+\gamma_1\epsilon_r}{\gamma_1\epsilon_z}\tilde{\epsilon}_z^\Delta\tilde{\epsilon}_z\\
\times \left[1+2\frac{\gamma_s\epsilon_c+\gamma_1\epsilon_r}{\gamma_1\epsilon_z}\left(\tilde{\epsilon}_z-\tilde{\epsilon}_z^\Delta\right)\right] \ .
\end{multline}
As expected, in the limit $\Delta_\text{S}\to \infty$, we recover the equations reported in the main text.

We remark also that the equations of $g$-factors and Rabi frequencies in Sec.~\ref{sec:SQD} and Sec.~\ref{sec:Elongated-QDs} are straightforwardly modified in the presence of a significant contribution of the SOHs by  substituting $\tilde{\kappa}\to \tilde{\kappa}_\text{S}$, $\tilde{\epsilon}_z\to \tilde{\epsilon}_z^\text{S}$, and $\tilde{\gamma}\to\tilde{\gamma}_\text{S}$, and by using the SOI velocities and masses defined in Eq.~\eqref{eq:pars_SOH}.

In Fig.~\ref{fig:pars_sohs} we  show the effect of the SOHs on the effective parameters, also including a comparison between the simple expressions reported in Eq.~\eqref{eq:pars_SOH} and the full expressions. We show the relevant quantities $m_{z,\varphi}= (1/m_*\pm 1/\delta m)^{-1}$,  $v_{z,\varphi}=v_-\pm v_+$ and the Zeeman parameters $\tilde{\epsilon}_z^\text{S}$ and $1-\tilde{\gamma}_\text{S}/\kappa$. 
We note that the Zeeman parameters and $v_z$ are not significantly affected by the SOHs, but the masses $m_{z,\varphi}$ and $v_\varphi$ are strongly renormalized by $\Delta_\text{S}$, especially in the presence of a large longitudinal strain $\epsilon_z$.

At low electric fields, the lower values of $v_\varphi$ due to the SOHs results in a smaller SOI-induced gap $\Delta=\hbar v_\varphi/R$, see Eq.~\eqref{eq:delta}. However, as argued in the text this gap becomes dominated by external electric field $E_x$ and already at weak values of $E_x$ it approaches $\omega_E\propto 1/\sqrt{m_\varphi}$, see Eq.~\eqref{eq:omegaE}. This energy gap becomes smaller because the value of $m_\varphi$ is  increased by the SOHs, however the gap remains in the meV range for the parameters examined. For example at $R=20$~nm, $\tau=10$~nm, and large strain ($R_3\to \infty$), $\omega_E$ is reduced by the multiplicative factor  $\sqrt{m_\varphi(\Delta_\text{S}\to \infty)/m_\varphi}\approx 0.8$, see Fig.~\ref{fig:pars_sohs}.

Strikingly, we find that the  SOHs can be beneficial to reach even higher values of spin-resonator coupling.
In fact, in the low electric field limit, the susceptibility of the spin qubit to the electric field is enhanced by the SOHs, because $\delta \underline{g}^{x,y}\propto \sin(\theta_E)$, see Eqs.~\eqref{eq:delta_g_app} and~\eqref{eq:delta_g_app_Ex}, and the angle $\theta_E=\arctan[eE_xR/\Delta\cos(\theta_z)]$ becomes larger by the reduced values of $\Delta$; the prefactor of these equations is also enhanced by the larger mass $m_\varphi$.
While these additional contributions are found to improve the performance of the qubit, to simplify the discussion, in the main text we do not examine them.

\section{Anisotropic corrections}

\label{sec:deviation_LK}

\begin{figure}
\centering
\includegraphics[width=0.5\textwidth]{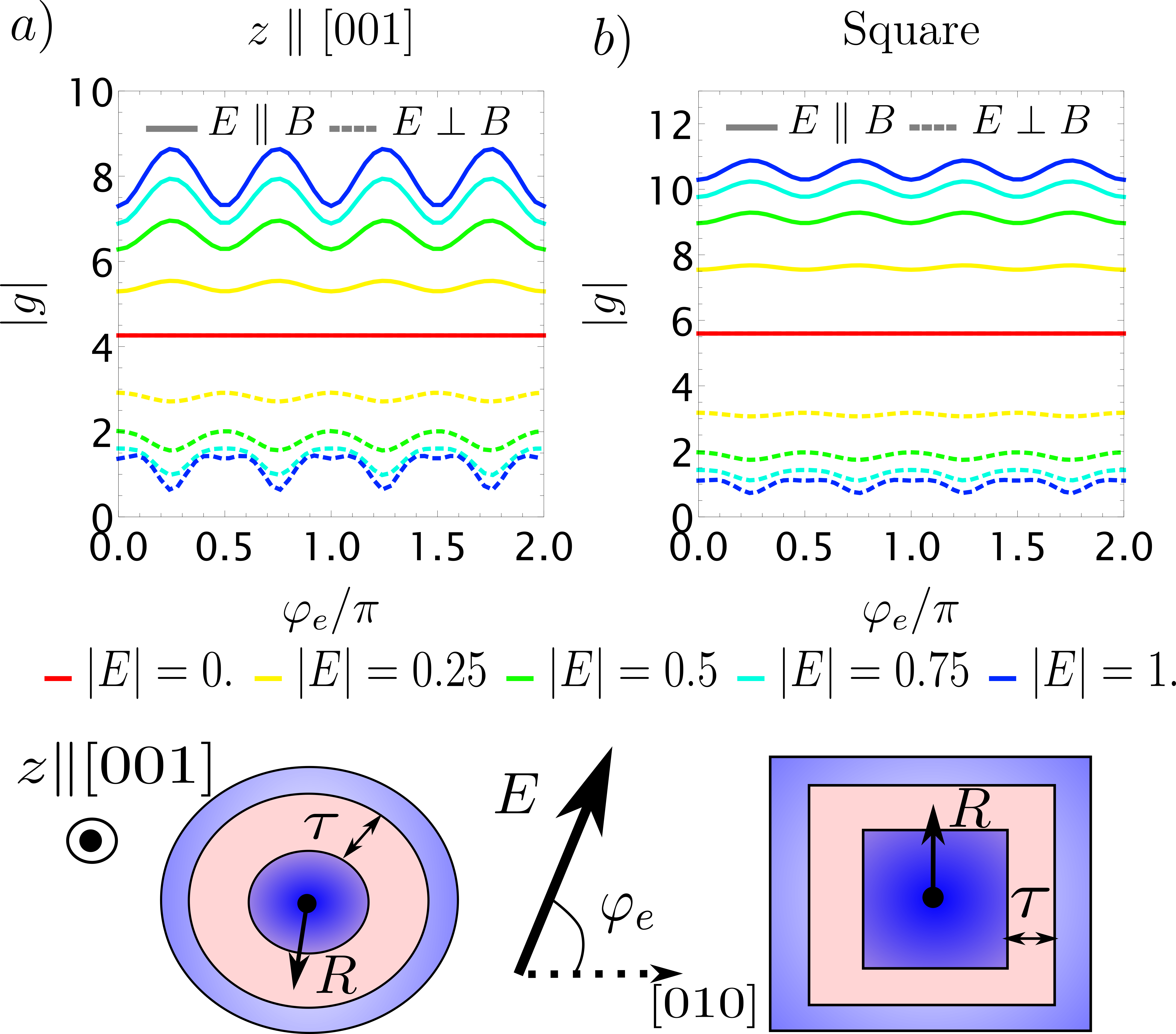}
\caption{\label{fig:gfact_SQD-ani}
Effect of the anisotropies on the $g$-factors of a hole spin qubit in a Ge CQW. We show with solid and dashed lines the total $|g|$-factor obtained when $\textbf{B}$ is parallel and perpendicular to $E$, respectively. We observe oscillations of $|g|$ as a function of the angle $\varphi_e$. These oscillations are comparable in panel a) and b), where we show a cylindrical quantum well grown along the $z\parallel [001]$ direction and a quantum well with square cross-section, respectively.
The setup  considered is shown at the bottom of the plots. In both cases, we consider $\epsilon_r=3\epsilon_z=\epsilon_c$, $\tau=R/2$, and $l_z=R$. The values of the electric field are given in unit of $\hbar\omega_\varphi/eR$.
 }
\end{figure}

We analyze here some additional properties of the system that go beyond the isotropic LK Hamiltonian in Eq.~\eqref{eq:LK-Hamiltonian}. 
In particular,  we study numerically the role of the anisotropies of the cubic lattice and variations from a circular cross-section. Because of these anisotropies, there are some energetically favoured directions where the hole is more likely to localize, and thus the electric field response becomes oscillatory depending on the direction of $\textbf{E}$.
By investigating separately cubic and cross-section anisotropies, we verify that these oscillations have a rather small amplitude and  our theory based on the isotropic LK Hamiltonian and circular cross-sections remains reasonably accurate  in more general cases.

The cubic anisotropies of the LK Hamiltonian are small in Ge because $(\gamma_3-\gamma_2)/\gamma_1\approx 0.1$, however because they break the rotation symmetry of Eq.~\eqref{eq:LK-Hamiltonian}, they induce qualitatively new terms in the effective Hamiltonian in Eq.~\eqref{eq:effective_H}. Because of these terms,  the total angular momentum is not conserved even without external fields. The amplitude and form of these terms depend on the growth direction of the well. Here, we restrict ourselves to the analysis of CQWs grown along the a main crystallographic axis, e.g. $z\parallel [001]$, and we show that the additional terms $\propto \gamma_3-\gamma_2$ only result in an additional small correction to the effective theory presented in Sec.~\ref{sec:SQD}. 

In Fig.~\ref{fig:gfact_SQD-ani}a), we show the effect of the cubic anisotropies of the LK Hamiltonian on the $g$-factor of a short quantum dot with $l_z=R$. We restrict our analysis here to the $g$-factor obtained for magnetic fields that are perpendicular to the quantum well; a detailed analysis of $g_{zz}$ for different growth directions is provided in~\cite{adelsberger2}. Qualitatively, we observe a similar behaviour as the one described in the main text. At $E=0$ the $g$-factor in $x\parallel [100]$ and $y\parallel[010]$ directions coincide, and at larger values of $E$, the $g$-factor is increased (decreased) when $\textbf{B}\parallel \textbf{E}$ ($\textbf{B}\perp \textbf{E}$). However, by rotating the electric field from the main  crystallographic axes by an angle $\varphi_e$, we observe that the LK anisotropies result in an additional oscillation with period $\pi/2$ superimposed to the isotropic response, see Fig.~\ref{fig:gfact_SQD}. We note that these oscillations have a larger amplitude and a less sinusoidal shape at larger values of $E$. We find similar oscillations on top of the isotropic response also in the driving terms $\delta g_{ij}^{x,y}$. 

Similar oscillations are found if the cross-section is not circular. For example, in Fig.~\ref{fig:gfact_SQD-ani}b), we show the $g$-factor of a short quantum dot defined in a thin Ge quantum well with a square cross-section. To investigate the role of cross-section anisotropy, we show here results obtained by using the isotropic LK Hamiltonian. We note that the $g$-factor in this case is qualitatively similar to the $g$-factor in Fig.~\ref{fig:gfact_SQD-ani}a), with only some small quantitative differences coming from the different cross-section. In particular, because of the 4-fold rotational symmetry of the square, the $g$-factor in this qubit is also oscillating as a function of $\varphi_e$ with period $\pi/2$ and the oscillations caused by the square cross-section have a small amplitude that increases at larger $E$.
Because these oscillations are rather small corrections to isotropic theory, we believe that our theory accurately describes a wide range of devices.

\bibliography{references}
\end{document}